\documentclass[cits]{PoS}

\usepackage[intlimits]{amsmath}
\usepackage{amssymb}
\usepackage{bbm}

\title{Review of lattice studies of resonances}

\ShortTitle{Review of lattice studies of resonances}

\author{\speaker{Daniel Mohler}\\
TRIUMF, 4004 Wesbrook Mall, Vancouver, BC V6T 2A3, Canada\\
Fermilab, PO BOX 500, Batavia, IL, 60510, USA \\
E-mail: \email{dmohler@fnal.gov}}

\abstract{I review recent progress in extracting resonance parameters using lattice field theory, with an emphasis on determining hadron resonances from lattice quantum chromodynamics. Until recently, the $\rho$-meson channel was the only one considered, while, during the last year, several resonant channels have been investigated for the first time. Recent lattice results for scattering phase shifts in resonant channels are presented. }

\FullConference{The 30th International Symposium on Lattice Field Theory\\
                 June 24  29,  2012\\
                 Cairns, Australia}

\begin{document}

\section{Introduction and methods}

Recent years have shown remarkable progress in lattice simulations of quantum chromodynamics (QCD). Dynamic simulations with multiple flavors of light quarks are now routinely performed and there are first calculations at physical pion masses. Properties of ground states can now, in many cases, be determined with small statistical errors and with fully controlled systematic uncertainties. At the same time new computational methods for smeared interpolating fields \cite{Peardon:2009gh,Morningstar:2011ka} and the use of the variational method \cite{Michael:1985ne,Luscher:1990ck,Blossier:2009kd} enable extraction of excited energy levels both in an unprecedented number and with unprecedented statistical accuracy \cite{Dudek:2010wm,Dudek:2011tt,Liu:2012ze}. However, most hadronic excitations are resonances and can decay strongly. Therefore the interpretation of the resulting data is straightforward only in the limit of very narrow states, which is often not the case in nature. 

\begin{table}[tbh]
\begin{center}
\begin{tabular}{cc|cc|cc}
\hline
hadron & $\Gamma$ [MeV] & hadron & $\Gamma$ [MeV] & hadron & $\Gamma$ [MeV]\\
\hline
$b_1(1235)$& $142\pm9$ & $K^\star(1410)$&$232\pm21$ & $D_0^\star(2400)$&$267\pm40$\\
$a_1(1260)$& $250-600$ & $K_0^\star(1430)$&$270\pm80$ & $D_1(2430)$&$384\pm^{130}_{110}$\\
\hline
\end{tabular}
\caption{Examples of light, strange-light and charm-light meson resonances. Values from the PDG compilation \cite{pdg12}.}
\label{example_hadrons}
\end{center}
\end{table}

Indeed, taking a look at the meson tables in the Particle Data Group (PDG) compilation \cite{pdg12}, resonances with a substantial hadronic width are more the norm than the exception. Table \ref{example_hadrons} lists examples for light, strange-light and charm-light mesons and their PDG values for the resonance parameters. As can be seen, there are many examples of resonances with sizable total widths, which should not be neglected. In these proceedings, progress in extracting resonance properties from lattice simulations is reviewed with an emphasis on recent simulations data rather than the associated theory.

In experiment, resonance properties are extracted using partial wave analysis. For the relatively simple case of elastic scattering, the scattering amplitudes $a_l$ is related to the scattering phase shift $\delta_l$ for the l-th partial wave:
\begin{align}
a_l&=\sin\delta_l\mathrm{e}^{i\delta_l}=\frac{\mathrm{e}^{2i\delta_l}-1}{2i}\;.
\end{align}
Near a single relativistic Breit-Wigner shaped resonance, one can then parametrize the scattering amplitude $a_l$ in terms of a resonance position $s_R=m_R^2$ and decay width $\Gamma$
\begin{align}
a_l&=\frac{-\sqrt{s}\Gamma(s)}{s-s_R+i\sqrt{s}\Gamma(s)}\;.
\end{align}

In a lattice QCD calculation in Euclidean space there is no direct access to the scattering amplitudes \cite{Maiani:1990ca}. However, as has been pointed out by L\"uscher \cite{Luscher:1986pf,Luscher:1990ux,Luscher:1991cf}, the phase shift of the continuum scattering amplitude in the elastic region can be determined from the discrete spectrum in a finite box. 

\subsection{The L\"uscher method}

\begin{figure}[tb]
\begin{center}
\includegraphics[width=6.0cm,clip]{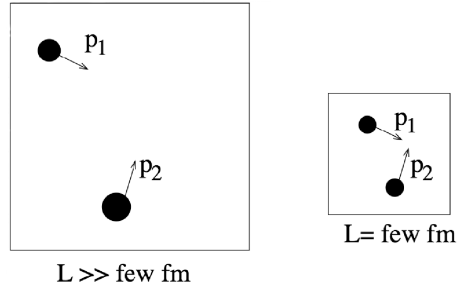}
\caption{Illustration of hadrons with momenta $p_1$ and $p_2$ in a finite volume. For a description please refer to the text. Figure taken from \cite{Prelovsek:2011nk}.}
\label{boxes}
\end{center}
\end{figure}

As mentioned, there is a relation between the phase shift of the continuum scattering amplitude in the elastic region and the discrete spectrum in finite volume \cite{Luscher:1986pf,Luscher:1990ux,Luscher:1991cf}. Figure \ref{boxes} illustrates the idea. In a large box with spatial volume $L^3$ and $L\gg1$~fm (depicted on the left-hand side) the energy of the two particle system is to a good approximation given by
\begin{align}
E=E(p_1)+E(p_2)\;,
\end{align}
where $E(p)=\sqrt(m^2+p^2)$ and $\vec p=\vec n(2\pi/L)$ in a relativistic simulation with periodic boundary conditions in space.
In a small box (depicted on the right-hand side) with $L\simeq 2\dots5$~fm the energy of the interacting system is noticeably shifted with regard to the non-interacting energy level. This energy shift is related by L\"uscher's formula to the elastic scattering phase-shift. Extracting resonance parameters from a lattice simulation using this relation therefore involves the following basic steps:
\begin{itemize}
\item [(1)] Extract the energy levels $E_n(L)$ in a finite box for one or more box sizes $L$.
\item [(2)] The L\"uscher formula relates this spectrum to the phase shift of the continuum scattering amplitude.
\item [(3)] Given the relevant phase shift data one can extract the resonance mass $m_R$ and the width $\Gamma_R$ or the coupling $g$ with some degree of modeling/approximation.
\end{itemize}

\begin{figure}[tb]
\begin{center}
\includegraphics[width=4.5cm,clip]{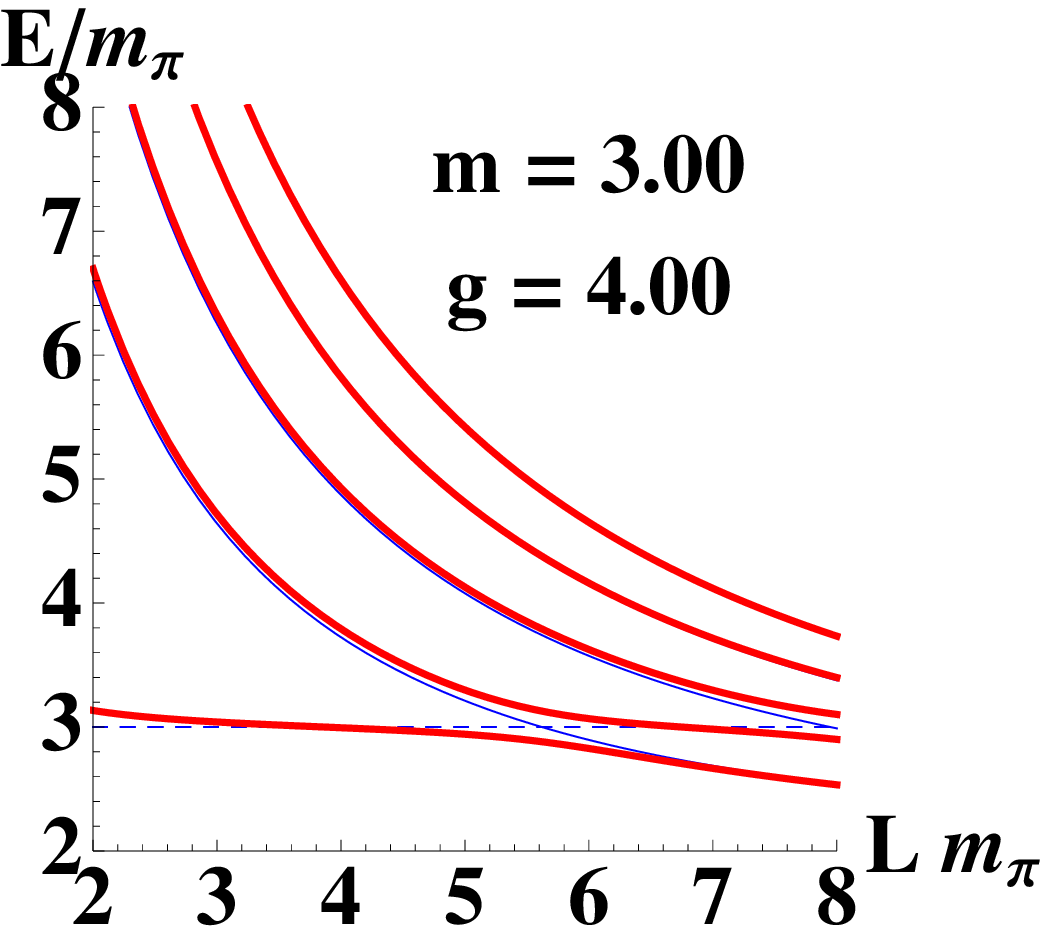}
\includegraphics[width=4.5cm,clip]{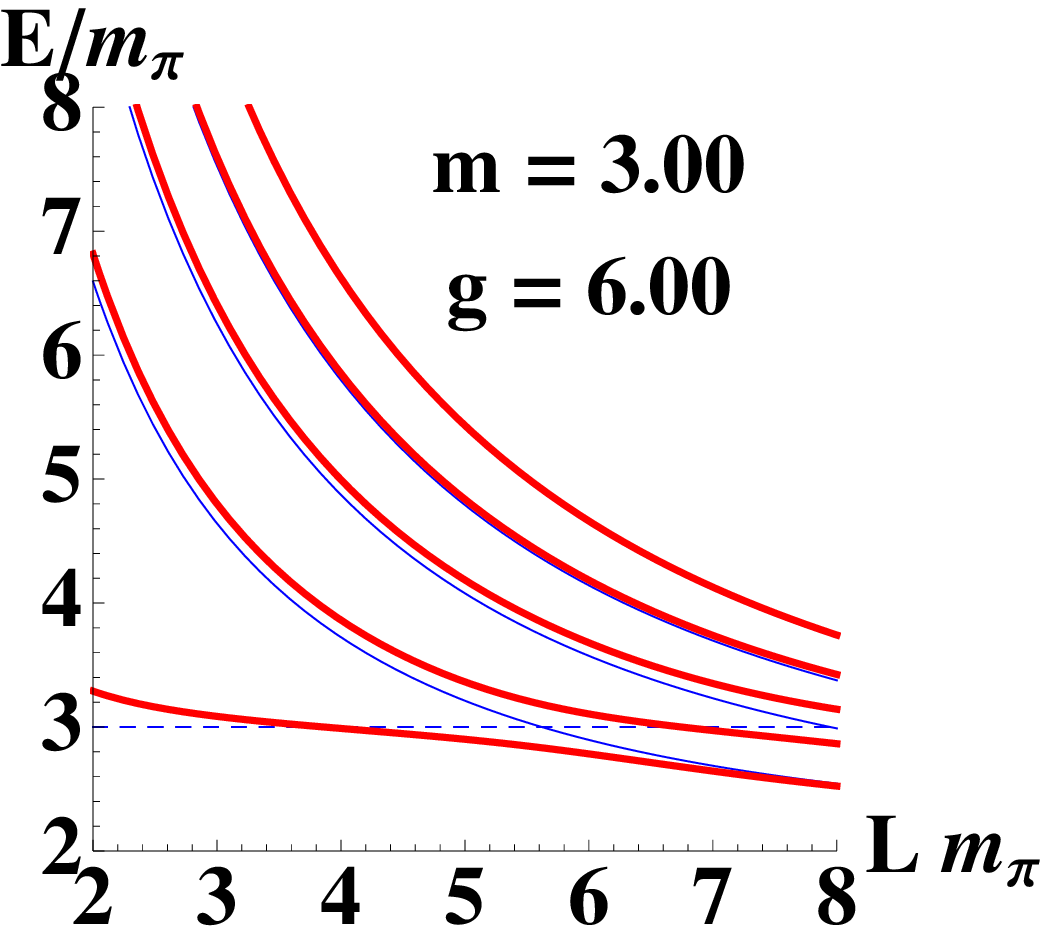}
\includegraphics[width=4.5cm,clip]{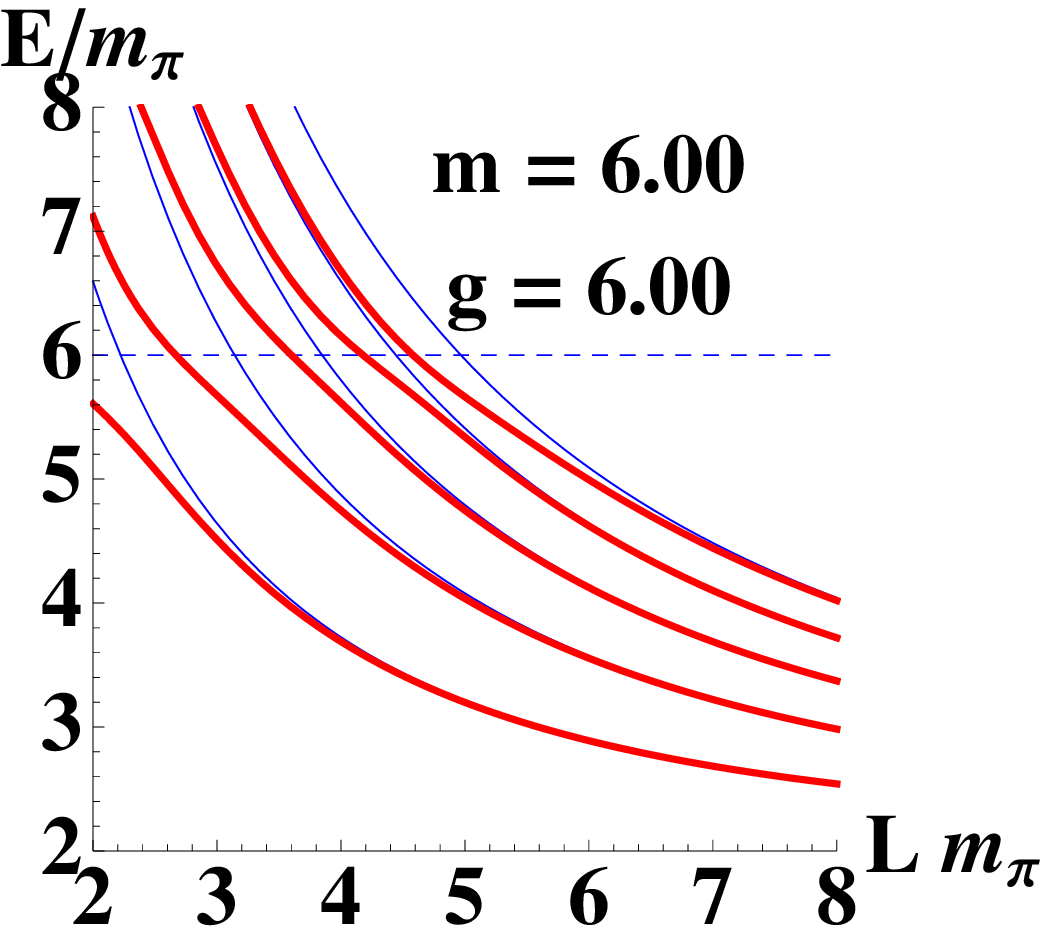}
\caption{Illustration of levels in a finite volume $L^3\times T$ for a $\rho$-like resonance. The resonance mass $m$ in units of the pion mass and the coupling $g$ are varied (assuming only elastic scattering and using the appropriate L\"uscher formula). The blue levels show the non-interacting levels and the blue dotted line indicates the resonance mass.}
\label{levels_illustration}
\end{center}
\end{figure}

To illustrate the finite volume spectrum, the few lowest energy levels for a $\rho$-meson-like system are plotted in Figure \ref{levels_illustration}. For a small coupling the typical avoided level crossing behavior is observed (left panel). With an increased coupling $g$ the energy shifts with respect to the non-interacting energy levels get more pronounced and the avoided level crossing gets more and more washed out (middle panel). The effect of increasing the mass at fixed coupling is also illustrated (right panel). From these illustrations it is clear that data for just a couple of energy levels in a single volume is of limited use. For a given lattice ensemble more data can be obtained by also considering moving frames.

While the original relation \cite{Luscher:1986pf,Luscher:1990ux,Luscher:1991cf} was limited to rest-frame calculation in multiple spatial volumes $L^3$, the corresponding L\"uscher formulae and interpolator constructions for equal mass hadrons $m_{h1}=m_{h2}$ in moving frames \cite{Rummukainen:1995vs,Kim:2005gf,Feng:2011ah,Dudek:2012gj} have been known for a while. During the last year the corresponding expressions for unequal mass hadrons $m_{h1}\ne m_{h2}$ in moving frames have been derived \cite{Fu:2011xz,Leskovec:2012gb,Doring:2012eu,Gockeler:2012yj}. It turns out that for the case of unequal mass hadrons, even and odd partial waves mix, creating an additional complication for the {\it ab initio } determination of phase shifts $\delta_l$ from lattice data. 

In addition to the above, there are some more recent ideas to extract resonances from lattice simulations \cite{Bernard:2008ax,Meissner:2010rq}. One of these, the so-called histogram method \cite{Bernard:2008ax} is illustrated briefly in the next section. For a description of the so-called correlator method please refer to \cite{Meissner:2010rq} directly.

\section{Methods in a toy model study}

In a recent publication, Giudice, McManus and Peardon \cite{Giudice:2012tg} tested both the L\"uscher method and the histogram method \cite{Bernard:2008ax} in the context of the O(4) non-linear sigma model, where precise data can be obtained: the authors of \cite{Giudice:2012tg} managed to accurately extract six energy levels for a range of $L/a=8,9,\dots,19$. For both methods, data was extracted in the elastic region (where the methods should be applicable) and in the inelastic region.

The histogram method \cite{Bernard:2008ax} can be summarized by the following procedure:

\begin{itemize}
\item [(1)] Determine the lowest few energy levels and interpolate to obtain a continuous function $E_n(L)$ in the interval $[L_0,L_M]$.
\item [(2)] Slice this interval $[L_0,L_M]$ into $M$ parts of length $\Delta L=\frac{L_M-L_0}{M}$.
\item [(3)] Slice the energy interval $[E_{min}, E_{max}]$ into bins of length $\Delta_E$.
\item [(4)] Make a histogram and normalize to get the distribution $W(E)$ or correspondingly $W(p)$.
\item [(5)] Subtract the non-interacting background $W_0(E)$ (or $W_0(p)$).
\item [(6)] In \cite{Bernard:2008ax} it is shown that close to a resonance one gets a Breit-Wigner shape
\begin{align}
W(p)-W_0(p)&\propto\frac{1}{[E(p)^2-M_r^2]^2+M_r^2\Gamma^2}\;, &W(p)=W(E)\frac{\partial E}{\partial p}\;.
\end{align}

\end{itemize}

\begin{figure}[tb]
\begin{center}
\includegraphics[clip, height=3.5cm]{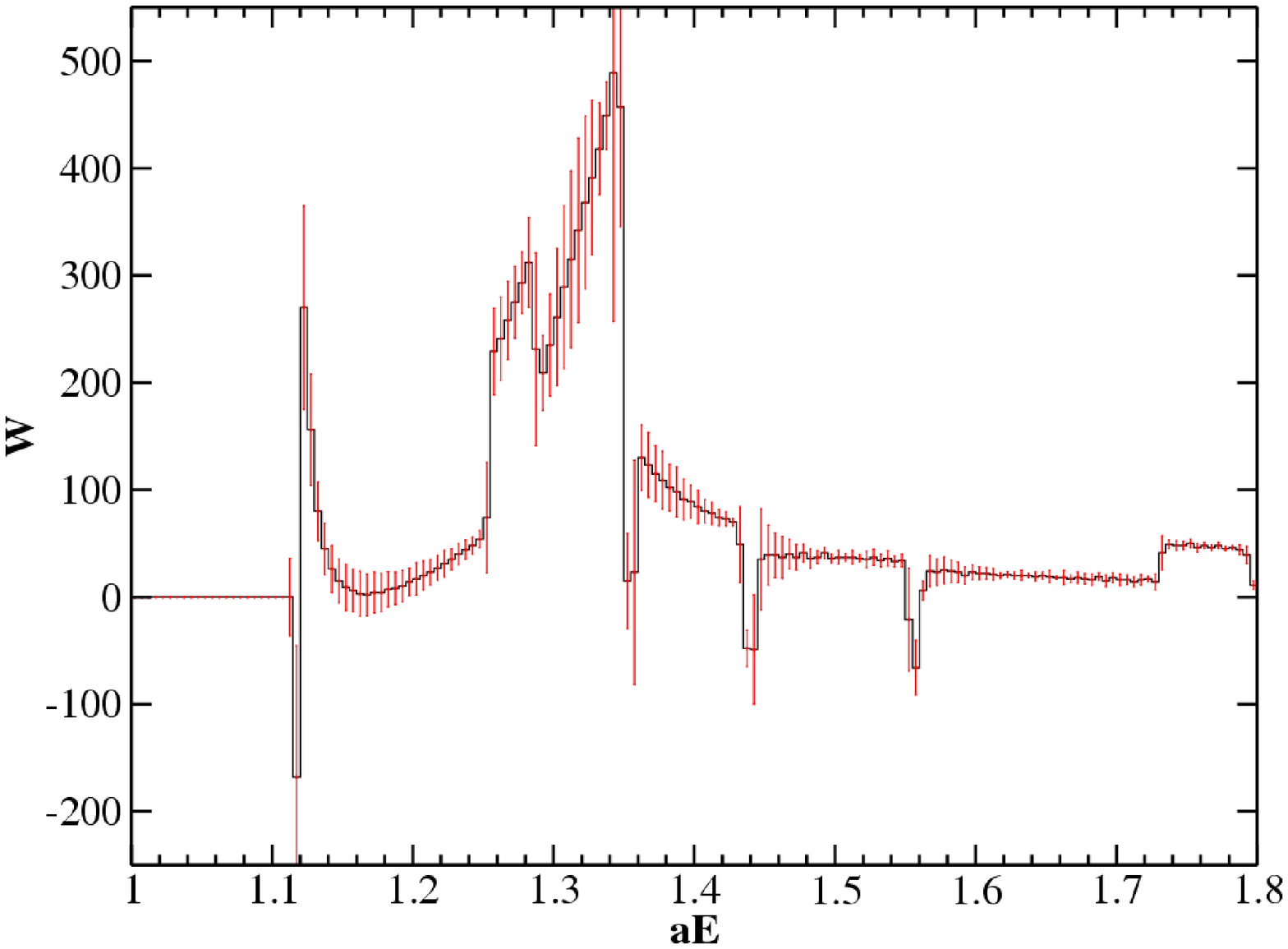}
\includegraphics[clip, height=3.5cm]{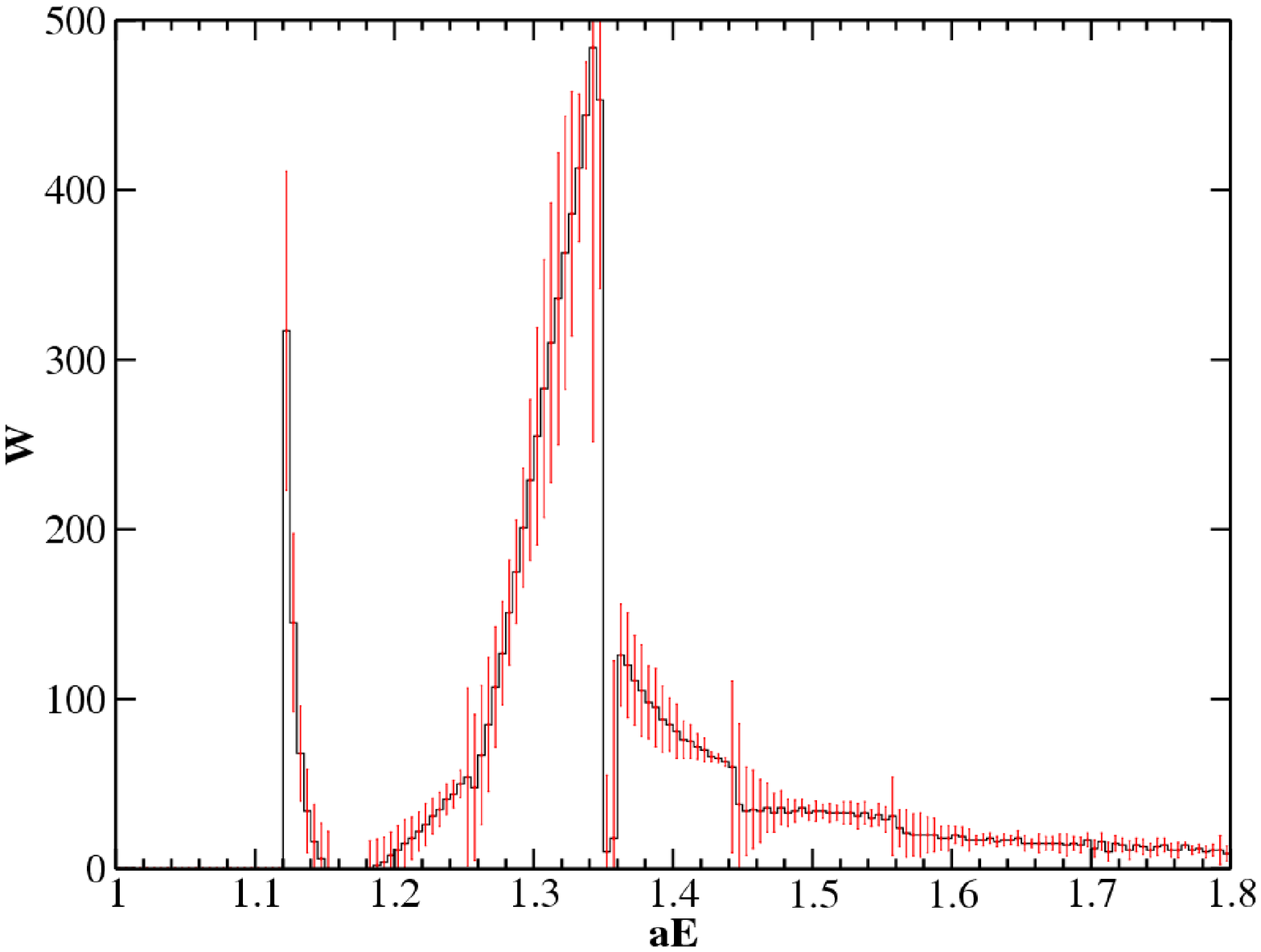}
\includegraphics[clip, height=3.5cm]{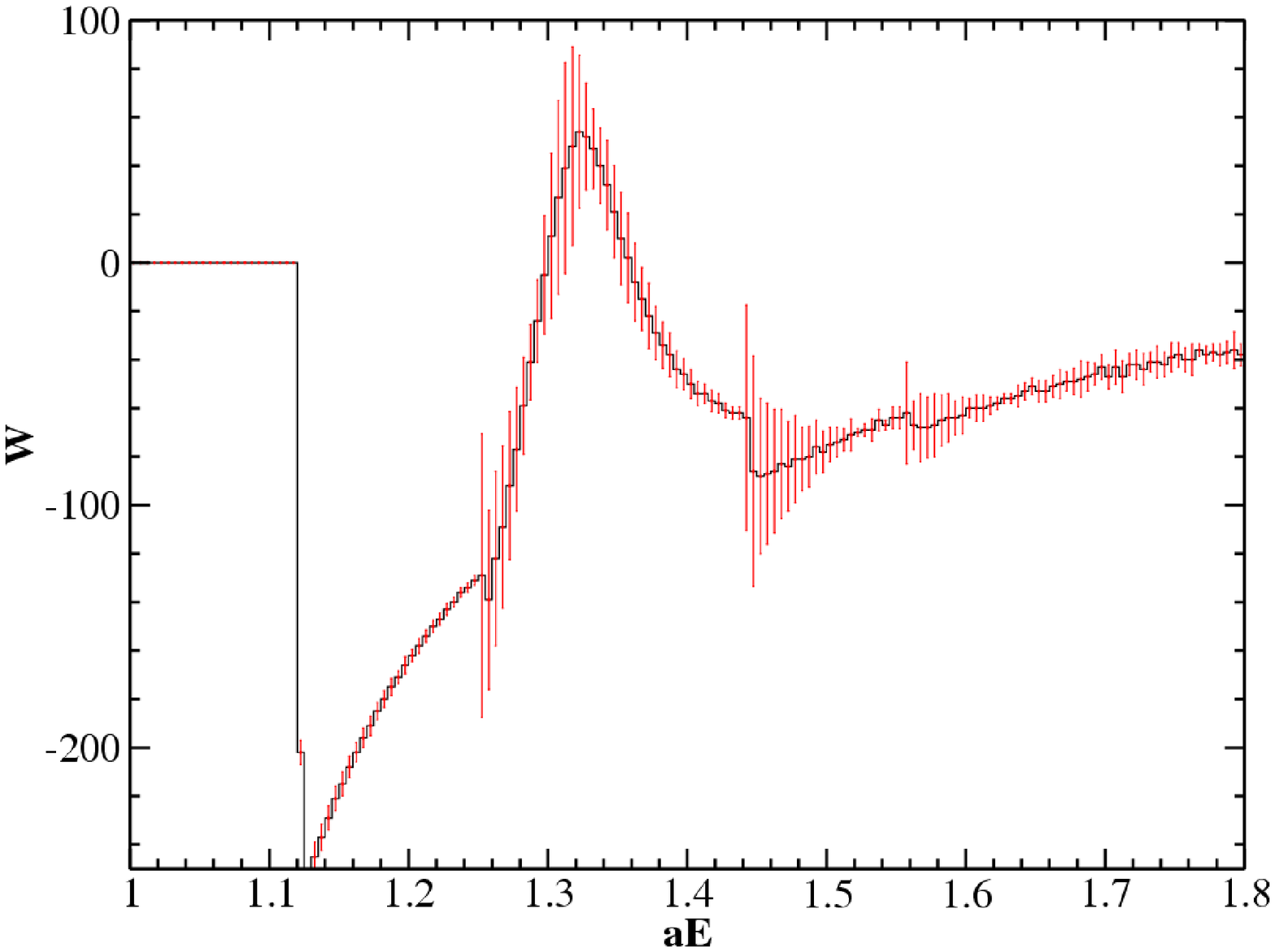}
\caption{Probability distribution $W-W_0$ for a narrow resonance with different strategies for the background subtraction. Figures taken from \cite{Giudice:2012tg}.}
\label{histogram_elastic}
\end{center}
\end{figure}

For further details please refer to \cite{Bernard:2008ax,Giudice:2012tg}. In practice, it turns out that the successful application of this procedure hinges on details of how to deal with the necessary background subtraction (step (5)). This behavior is illustrated in Figure \ref{histogram_elastic}, where a narrow resonance was investigated with different strategies for the background subtraction. The figure also illustrates that, in the context of this toy model, a narrow resonance can be extracted reliably once the background has been subtracted correctly. Unfortunately this was only achieved after omitting some of the data from the analysis. 

\begin{figure}[bt]
\begin{center}
\includegraphics[clip, height=4.5cm]{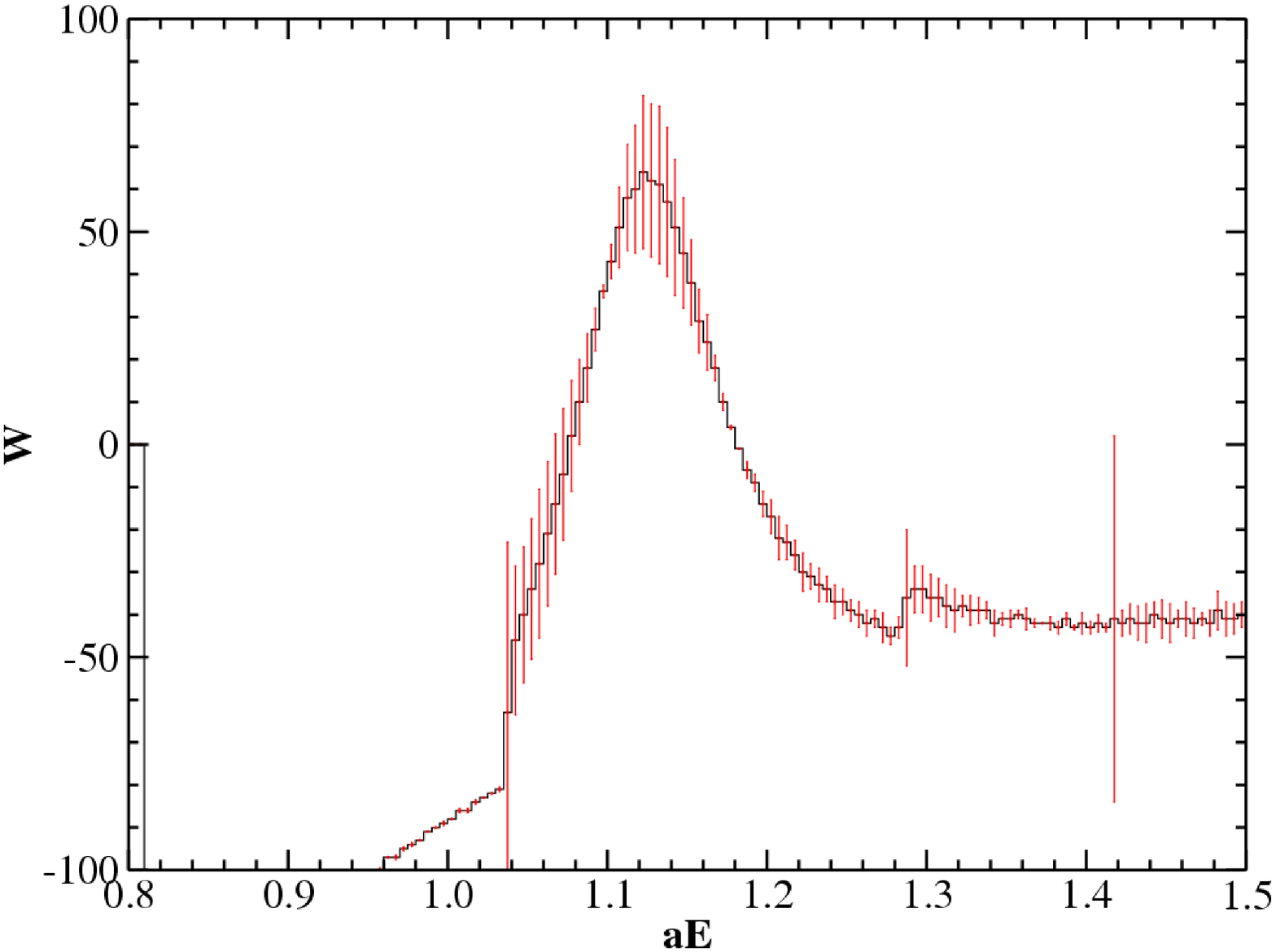}
\caption{Final results for a resonance in the inelastic regime. In this case, the relation to resonance parameters is unclear. Figure taken from \cite{Giudice:2012tg}.}
\label{hist_inel}
\end{center}
\end{figure}

The authors also apply the histogram method in the inelastic case, where there is no theoretical support for its applicability. Figure \ref{hist_inel} shows that nevertheless some sort of resonance shape emerges.

\begin{figure}[tb]
\begin{center}
\includegraphics[clip, height=3.5cm]{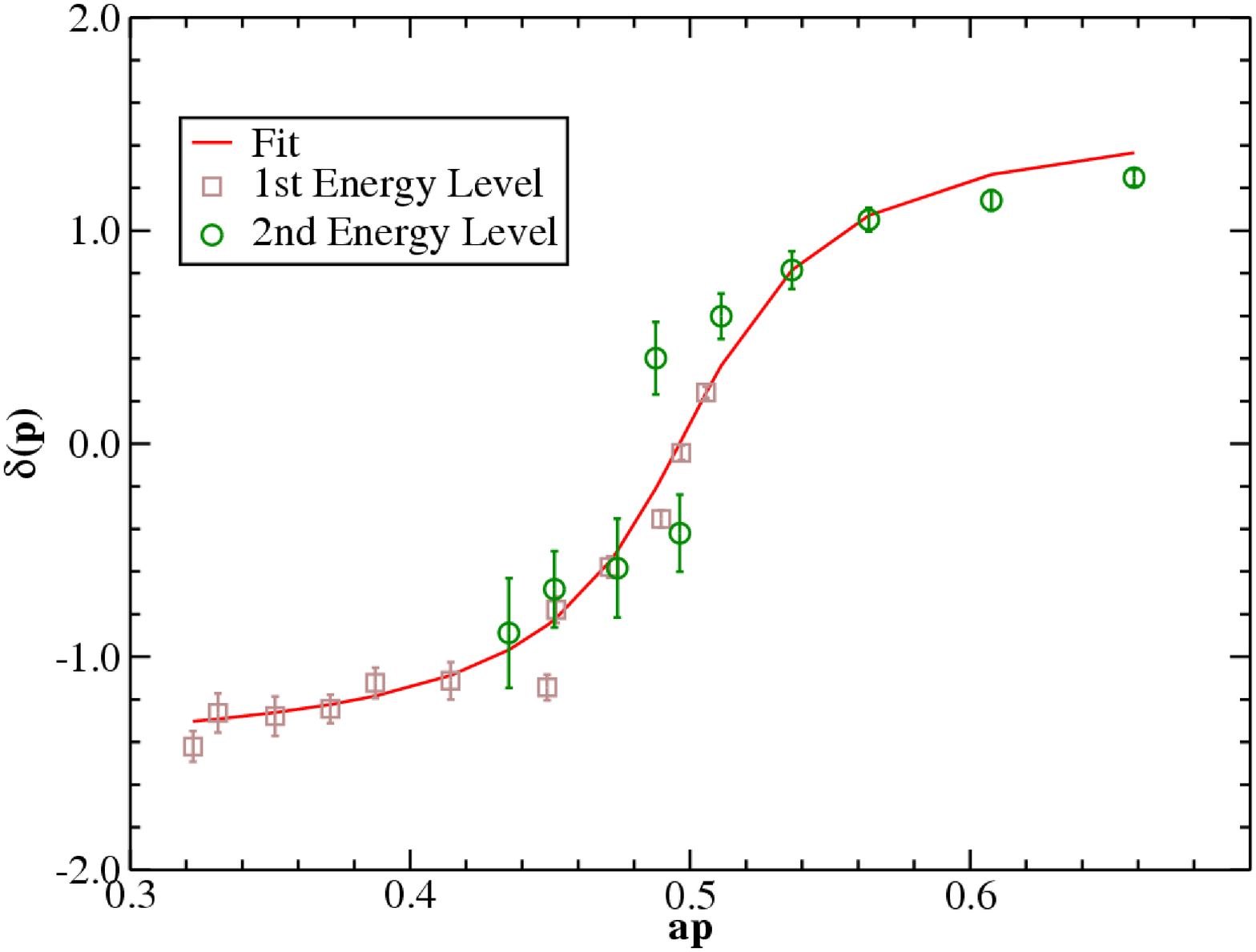}
\includegraphics[clip, height=3.5cm]{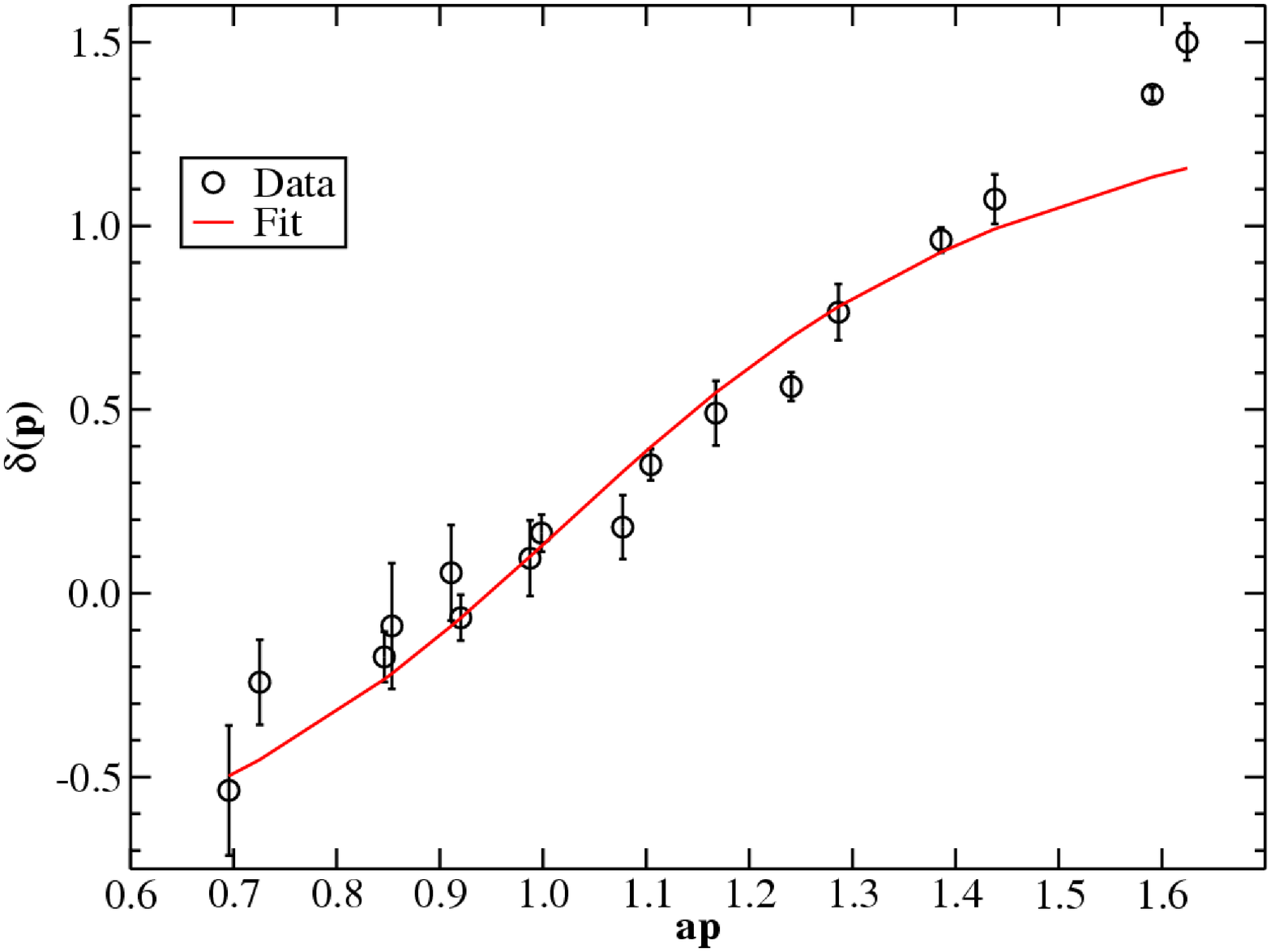}
\includegraphics[clip, height=3.5cm]{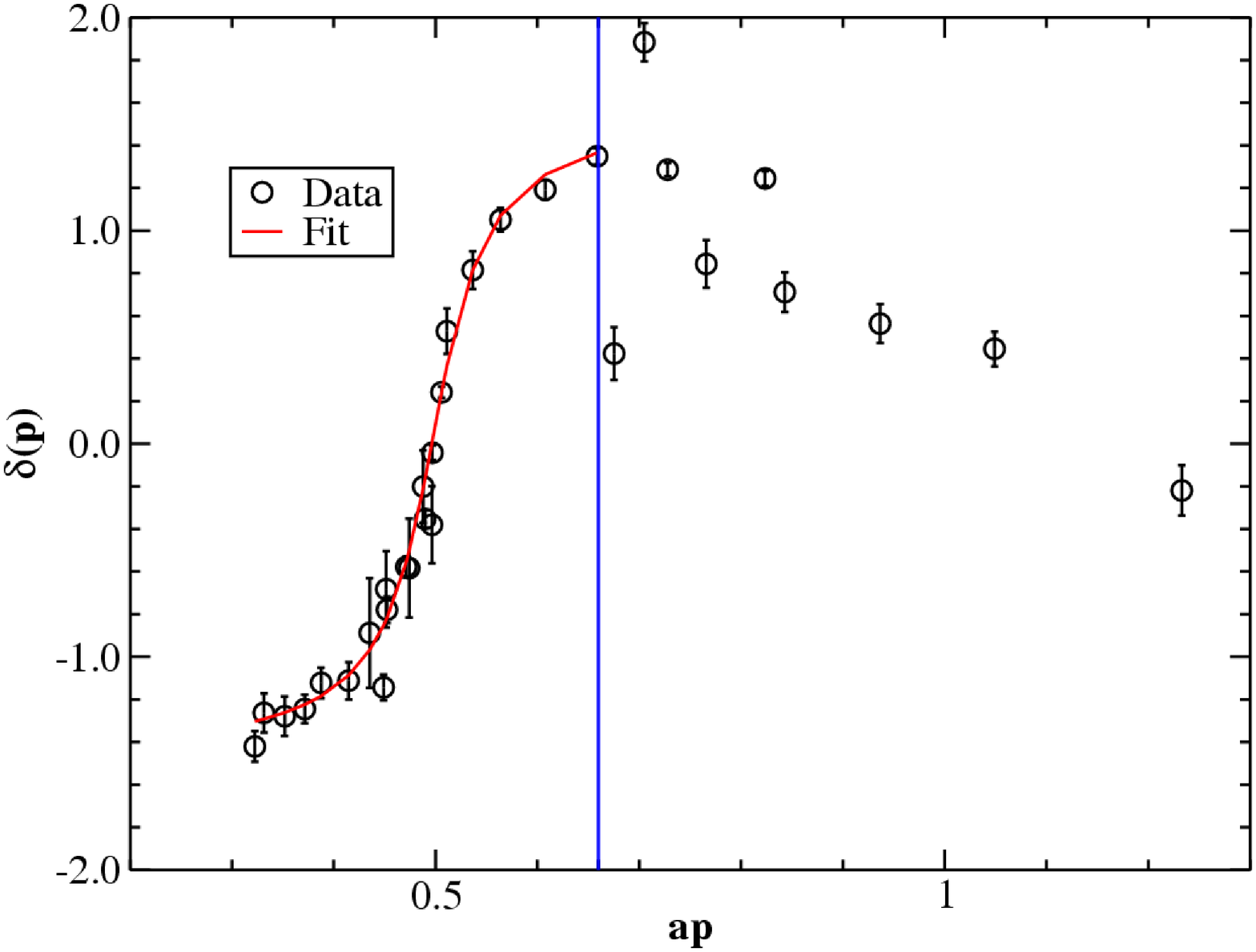}
\caption{Phase shift data obtained from the L\"uscher method.The blue line in the right plot shows the position of the inelastic threshold. For a description please refer to the text. Figures taken from \cite{Giudice:2012tg}.}
\label{luescher_all}
\end{center}
\end{figure}

Let us now turn to the L\"uscher method \cite{Luscher:1986pf,Luscher:1990ux,Luscher:1991cf} described in the previous section. The left and center panels in Figure \ref{luescher_all} show the results for a narrow and wide resonance. In both cases a clear resonance shape can be identified, although the resonance parameters can be determined much more accurately for a narrow resonance. In addition the authors also investigated what happens in the toy model when the method is applied above inelastic threshold. This can be seen in the right panel of Figure \ref{luescher_all}, where a narrow resonance is seen below threshold and the data becomes nonsensical above inelastic threshold. This illustrates that care needs to be taken when using the L\"uscher method for the extraction of resonance parameters.

\begin{table}[tbh]
\begin{center}
\begin{tabular}{ccccc}
\hline
&\multicolumn{2}{c}{L\"uscher}&\multicolumn{2}{c}{histogram}\\
parameter set & $aM_\sigma$ & $a\Gamma_\sigma$ & $aM_\sigma$ & $a\Gamma_\sigma$\\
\hline
set A & 1.35(2) & 0.115(8) & 1.33(5) & 0.10(5)\\
set B & 2.03(2) & 0.35(2)  & 2.01(2) & 0.35(10)\\
set C & 3.1(7)  & 1.2(5)   & - & - \\
\hline
\end{tabular}
\label{results_compare_table}
\caption{Comparison of resonance mass $M$ and width $\Gamma$ obtained with the L\"uscher and histogram methods for narrow (set A), medium (set B) and broad (set C) resonances \cite{Giudice:2012tg}.}
\end{center}
\end{table}

To compare how well the two methods fare, Table \ref{results_compare_table} lists the resonance parameters obtained with the help of the two methods. While the results from the two methods agree within their statistical uncertainty, the L\"uscher method leads to smaller statistical uncertainties and can be used to successfully extract broader resonances.

\section{QCD resonances}

\begin{figure}[tb]
\begin{center}
\includegraphics[clip, height=1.2cm]{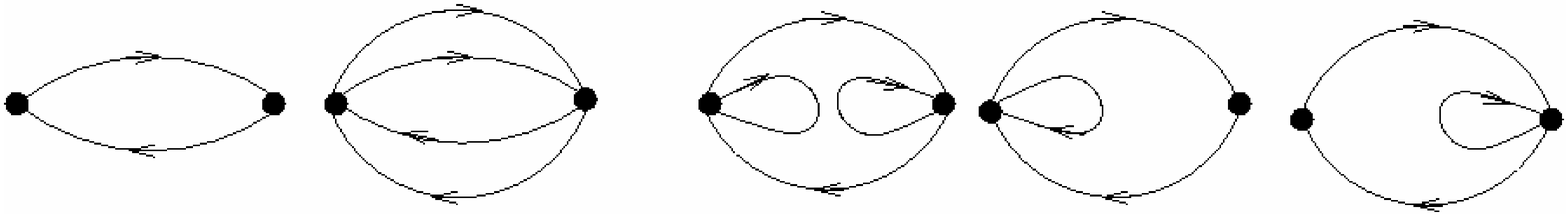}
\caption{Example of the Wick contractions arising in a meson-meson channel with $\bar{q}q$ and meson-meson interpolator basis.}
\label{contractions}
\end{center}
\end{figure}

Let us now turn our attention to the calculation of hadron resonances in QCD. These calculations are computationally demanding and very challenging from a technical point of view. It is observed in many calculations that $\bar{q}q$ operators conventionally used for the study of hadron excitations couple very weakly to multi-hadron states \cite{McNeile:2002fh,Engel:2010my,Bulava:2010yg,Dudek:2010wm}. A similar observation was made in string breaking studies \cite{Pennanen:2000yk,Bernard:2001tz}. This necessitates the inclusion of hadron-hadron interpolators. Figure \ref{contractions} shows an example of quark diagrams for a typical meson-meson channel. In particular some of the quark lines contain backtracking quark lines which are very expensive to calculate as they require the use of all-to-all propagator techniques. A particularly promising technique is the distillation method \cite{Peardon:2009gh,Morningstar:2011ka}.

\subsection{The $\rho$ meson: A benchmark calculation}

In experiment, the $\rho$ meson is seen as a p-wave resonance in $\pi\pi$-scattering. In many ways the $\rho$ meson is an ideal candidate for a benchmark calculation. While the physical $\rho$ can also decay into four pions, there is a large span of (unphysical) pion masses where the decay into two pions is the only possible decay. Furthermore, the resonance is of medium width (much smaller than its mass) and well isolated from other resonances. The signal in this channel is of a good quality even with a moderate number of gauge configurations. Naturally this was the first QCD resonance where L\"uscher's method has been applied to extract the mass and width of the state \cite{Aoki:2007rd}. In the meantime several groups have successfully demonstrated the feasibility of determining resonance parameters of the $\rho$ meson in a QCD calculation \cite{Feng:2010es,Aoki:2011yj,Lang:2011mn,Pelissier:2012pi,Gockeler:2008kc,Frison:2010ws}. In this section the current state of affairs is illustrated.

\begin{figure}[b]
\begin{center}
\includegraphics[clip, height=3.5cm]{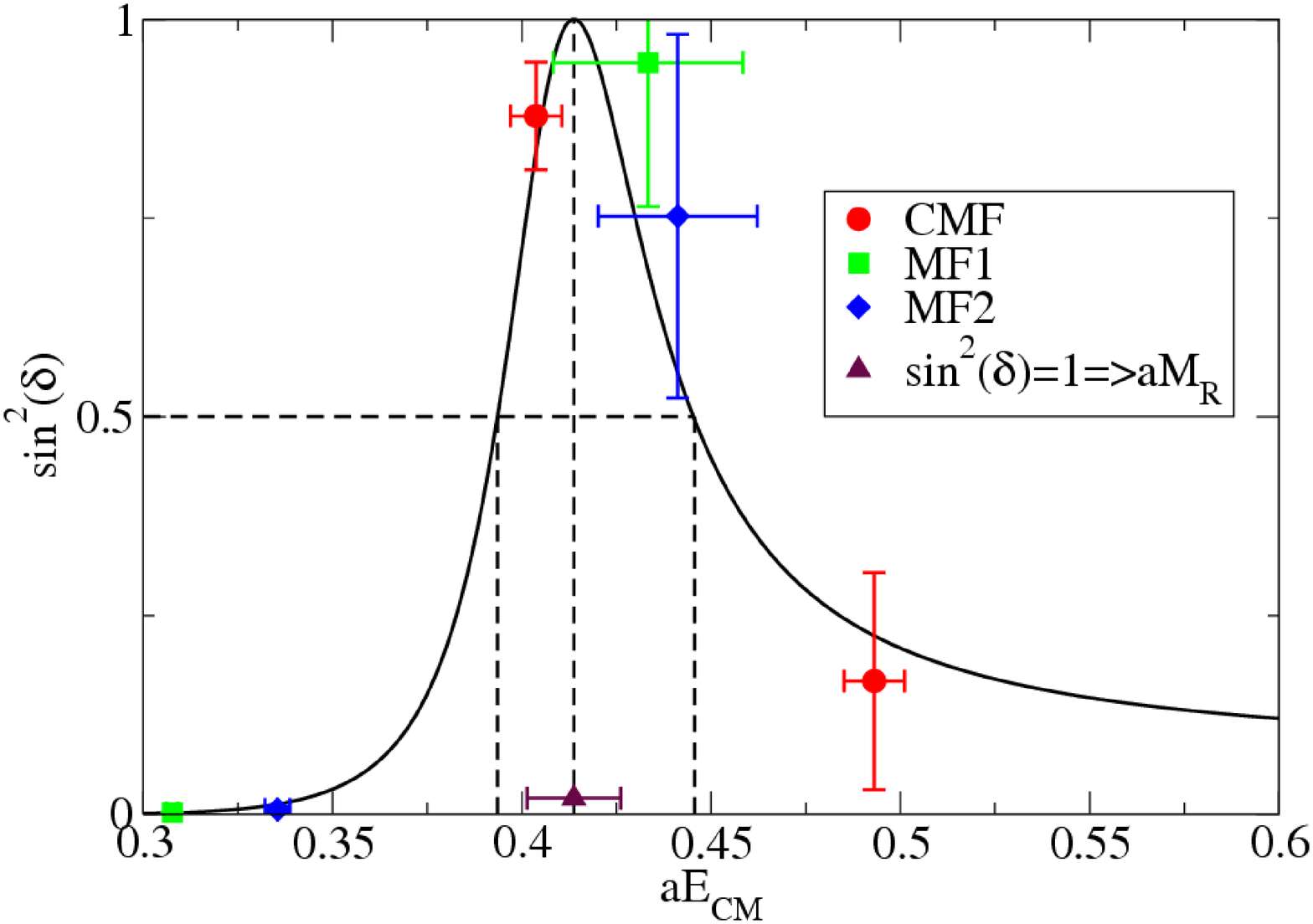}
\includegraphics[clip, height=3.5cm]{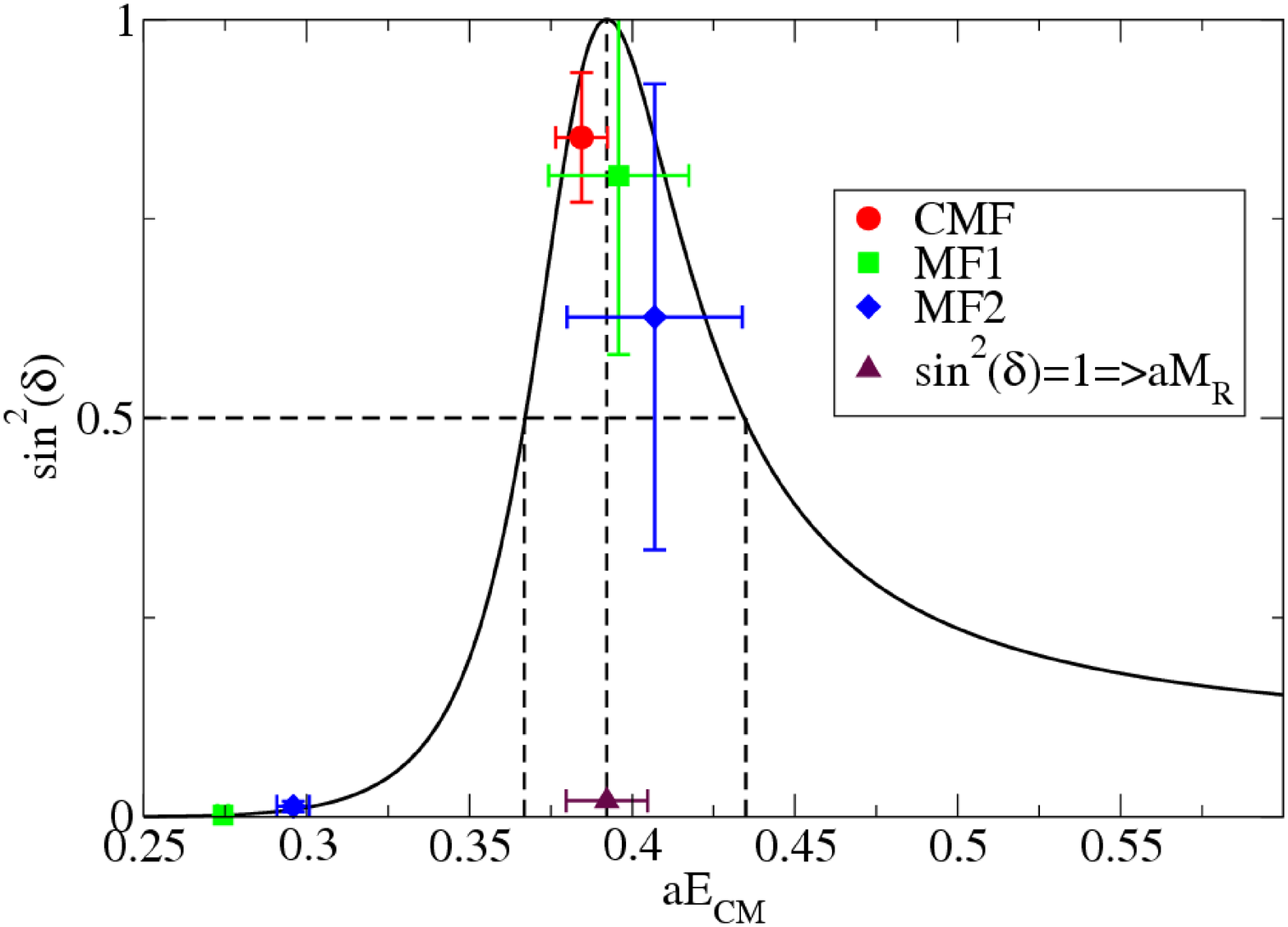}\\
\includegraphics[clip, height=3.5cm]{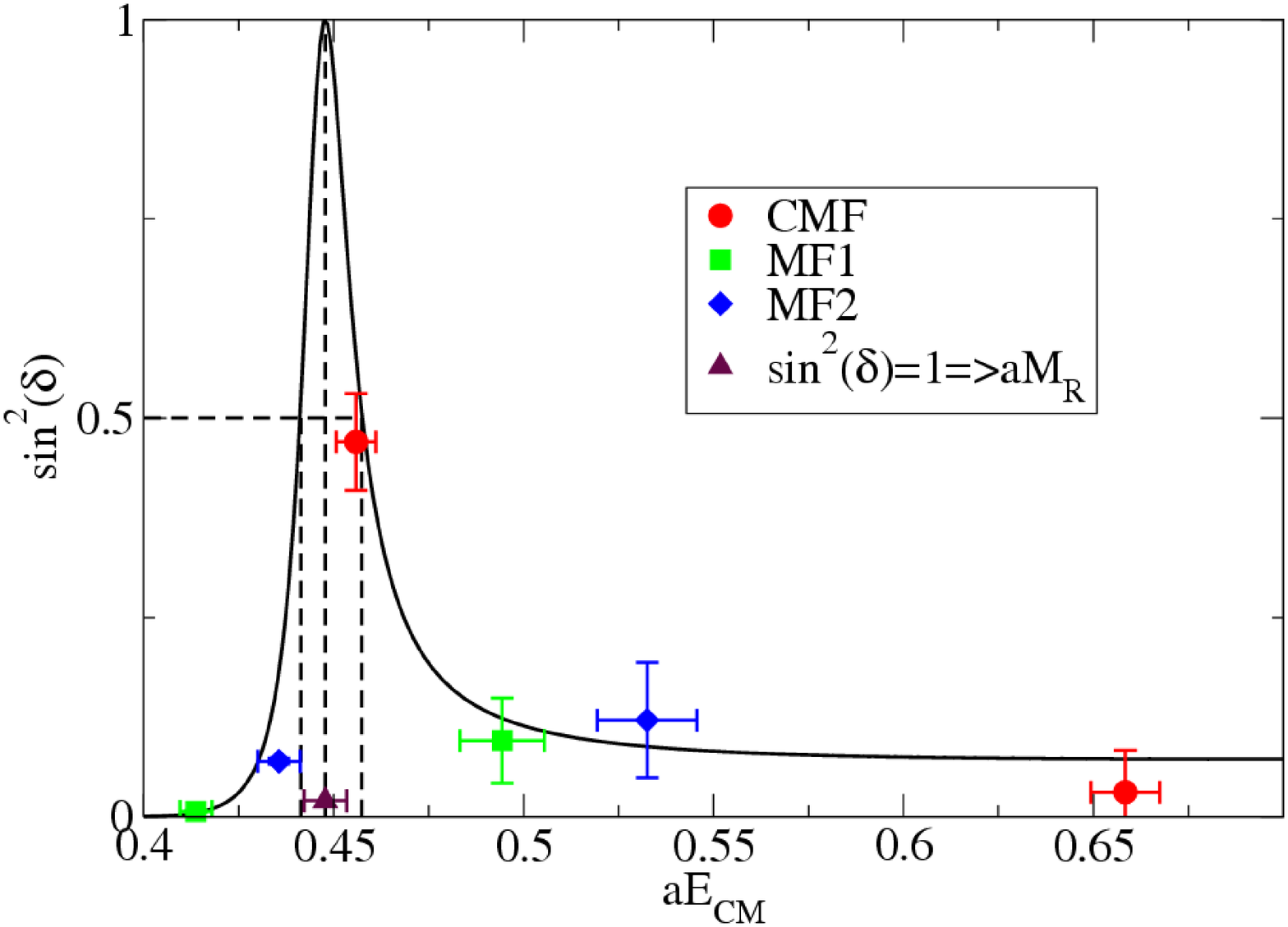}
\includegraphics[clip, height=3.5cm]{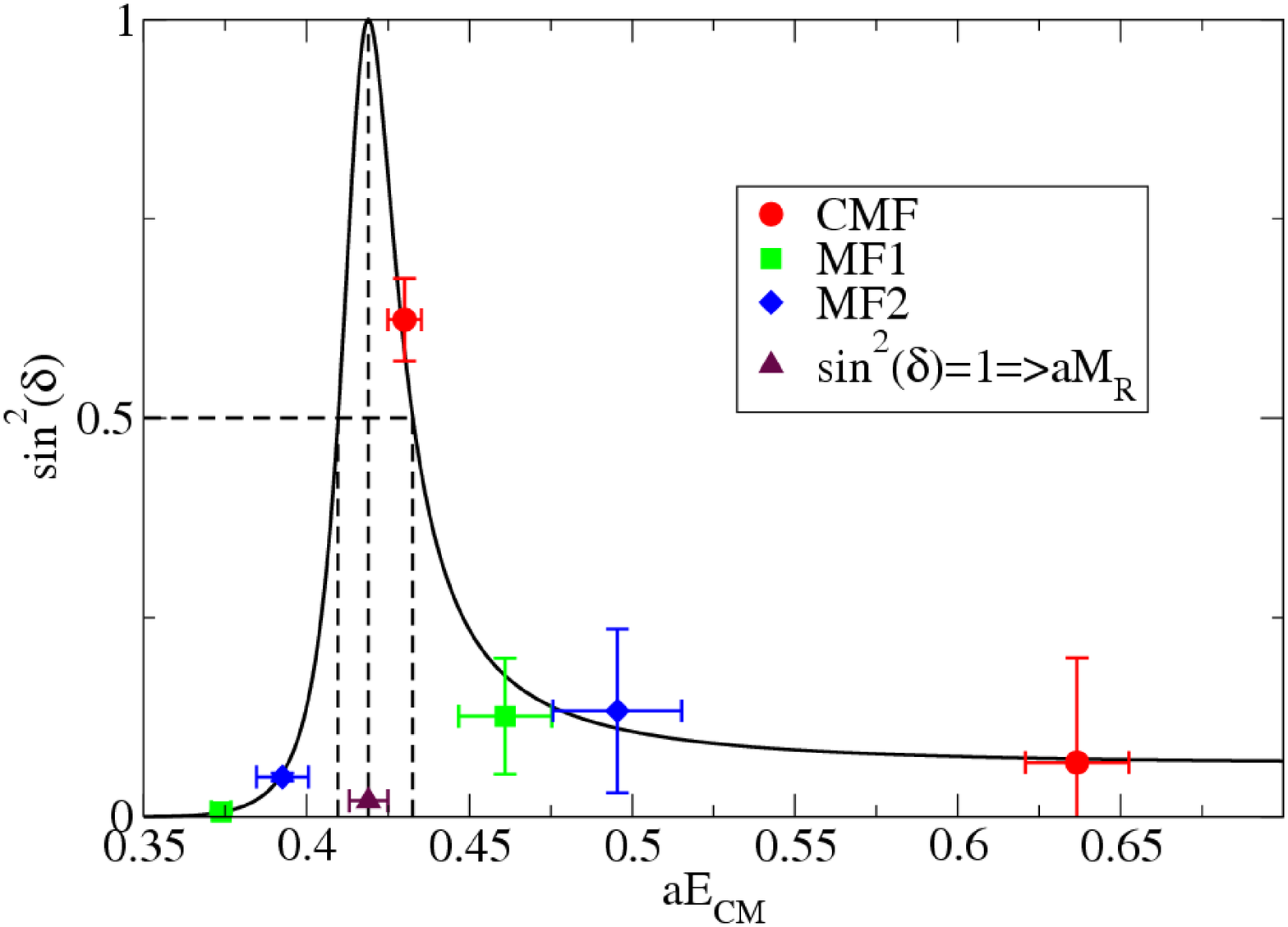}\\
\caption{Phase shift results for the $\rho$ meson resonance calculated by the ETM Collaboration \cite{Feng:2010es}. The four panels show results for gauge ensembles with four different pion masses $m_\pi\approx 290, 330, 420, 480$ MeV.}
\label{rhodata_etmc}
\end{center}
\end{figure}

Figure \ref{rhodata_etmc} shows the results obtained by the ETM Collaboration \cite{Feng:2010es}. They use four ensembles with the same lattice spacing and different light valence quarks and three momentum frames. This enables them to extract phase shifts for each ensemble and to perform an extrapolation to the chiral limit (assuming the coupling $g_{\rho\pi\pi}$ to be mass independent), while all of their data is extracted in the elastic regime where the framework is applicable.

\begin{figure}[tb]
\begin{center}
\includegraphics[clip, height=4.5cm]{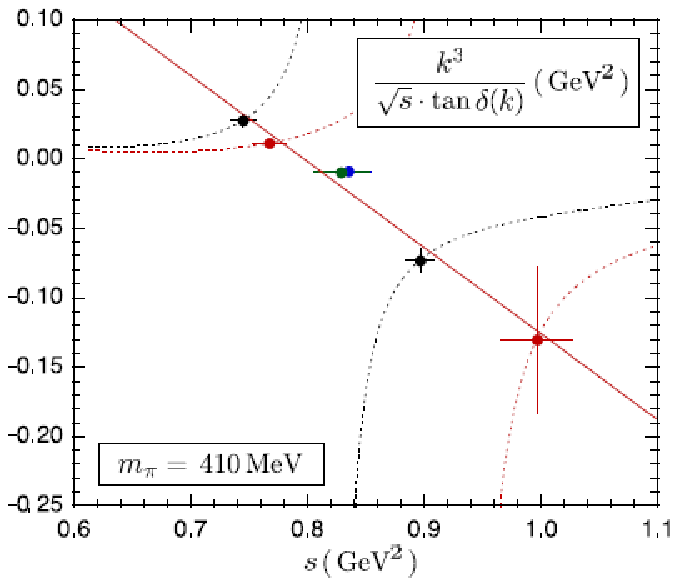}
\includegraphics[clip, height=4.5cm]{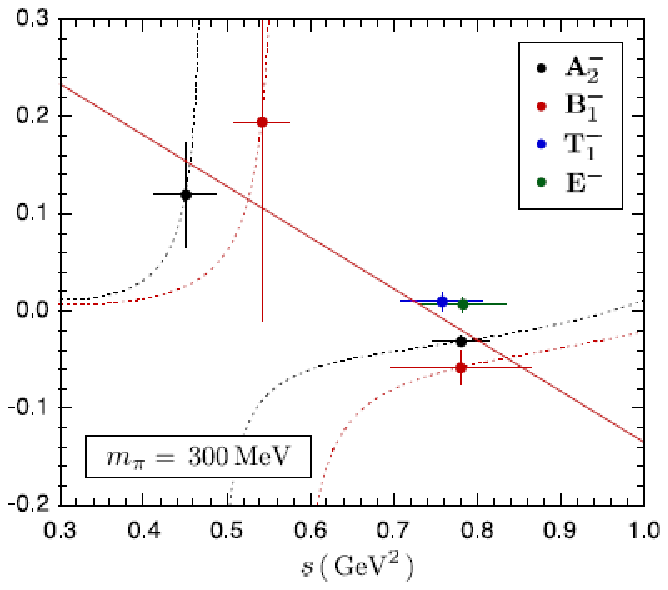}\\
\caption{Phase shift results for the $\rho$ meson resonance calculated by the PACS-CS collaboration \cite{Aoki:2011yj}. The pion masses used in the simulation are indicated in the panels.}
\label{rhodata_pacs-cs}
\end{center}
\end{figure}

Figure \ref{rhodata_pacs-cs} shows similar results from the PACS-CS collaboration using two ensembles and three momentum frames \cite{Aoki:2011yj}. Again, all data has been obtained in the elastic region.

\begin{figure}[tb]
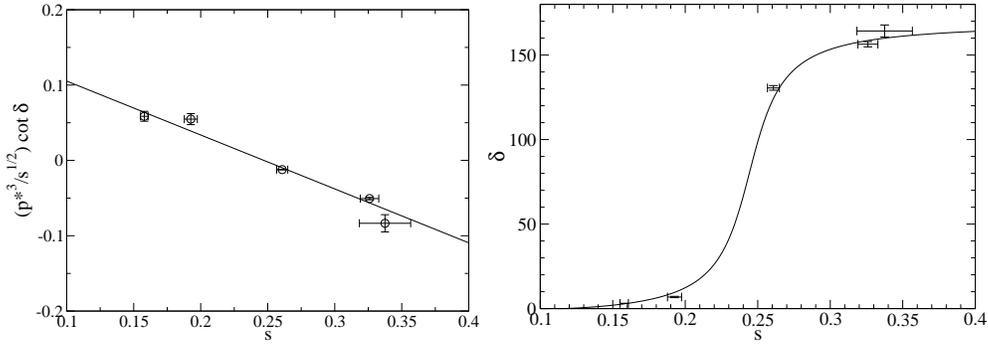

\begin{center}
\includegraphics[clip, height=4.5cm]{BW_xy_fits.eps}
\includegraphics[clip, height=4.5cm]{phaseshift_and_fit.eps}\\
\caption{Phase shift results for the $\rho$ meson from \cite{Lang:2011mn}.}
\label{rhodata_lmpv}
\end{center}
\end{figure}

Figure \ref{rhodata_lmpv} shows results by Lang {\it et al.} \cite{Lang:2011mn} from a single ensemble, demonstrating that a very good statistical accuracy can be achieved using the distillation technique \cite{Peardon:2009gh,Morningstar:2011ka}. All these results were presented in parallel talks at Lattice 2011.

\begin{figure}[tb]
\begin{center}
\includegraphics[clip, height=4.5cm]{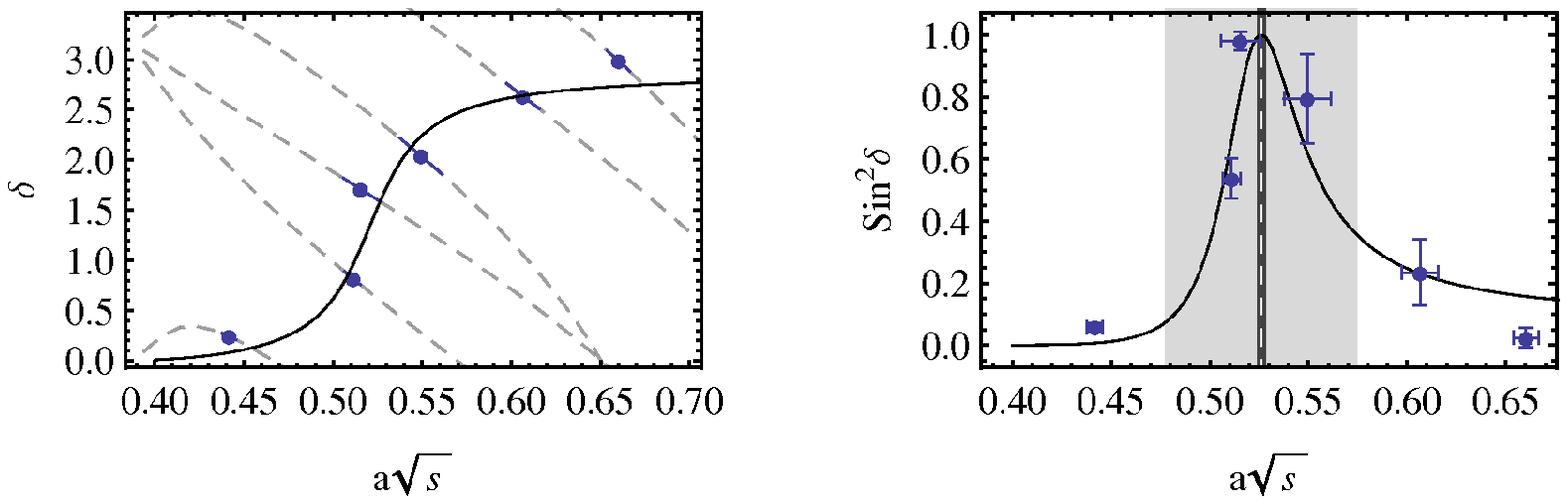}
\caption{Phase shift results from \cite{Pelissier:2012pi}, where asymmetric lattices were used to get phase shift points in the whole resonance region.}
\label{rho_pelissier}
\end{center}
\end{figure}

More recently, Pelissier and Alexandru \cite{Pelissier:2012pi} presented results from  a dynamical simulation on asymmetric lattices \cite{Feng:2004ua,Li:2003jn}, which followed an exploratory quenched simulation \cite{Pelissier:2011ib}. They generated three ensembles with $N_f=2$ nHYP-smeared clover fermions at a lattice spacing of $a=0.1225(7)$fm and  with a pion mass of $m_\pi=304(2)$MeV. Their lattice sizes are $24^2\times L_z\times48$ with $L_z=24,32,48$. To extract the two lowest states they use a variational basis with one $\bar{q}q$ and one meson-meson interpolating field. These results are plotted in Figure \ref{rho_pelissier}.

\begin{figure}[tb]
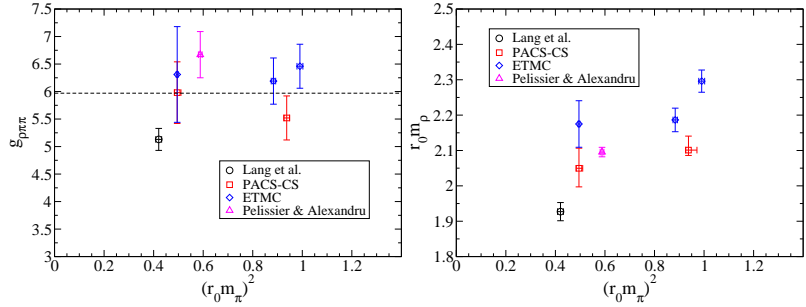

\begin{center}
\includegraphics[clip, height=4.0cm]{g_rho_compare.eps}
\includegraphics[clip, height=4.0cm]{m_rho_compare.eps}\\
\caption{Comparison of recent lattice results for the $\rho$ meson resonance mass (right panel) and coupling (left panel).}
\label{rho_compare}
\end{center}
\end{figure}

To compare all results for the resonance mass $m_R$ and the coupling $g_{\rho\pi\pi}$ it is best to plot the results in dimensionless units by multiplying the masses with the Sommer scale \cite{Sommer:1993ce} $r_0$. To this end the values for $r_0/a$ for each ensemble are used. The results are shown in Figure \ref{rho_compare} and the errors plotted encompass the statistical uncertainty and the uncertainty in the Sommer scale. Plotted in this scale-independent way all available results agree with each other fairly well, although the values obtained by Lang {\it et al.} for both the mass and the coupling are somewhat lower than the rest. This highlights the need to address various systematic uncertainties in more detail in future simulations.

\begin{figure}[tb]
\begin{center}
\includegraphics[clip, height=4.5cm]{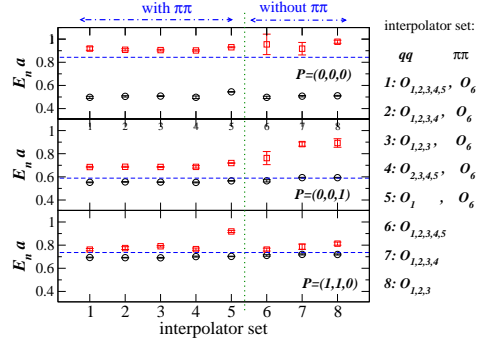}
\caption{Interpolator set dependence for the energy levels used to extract the $\rho$ resonance in \cite{Lang:2011mn}.}
\label{setdep}
\end{center}
\end{figure}

In this regard, it is instructive to take a look at the interpolator dependence of the results from \cite{Lang:2011mn} which are shown in Figure \ref{setdep}. On the left side of the figure the energy levels resulting from several choices including both pion-pion and $\bar{q}q$ interpolating fields are shown. Here, a basis of just two interpolating fields turns out to give rather unreliable results when compared to several choices of a bigger basis. Notice that all fits had an acceptable $\chi^2/d.o.f.$. On the right side of the plot choices of basis without a pion-pion interpolator are shown. In all cases the errors are larger than for the mixed basis, especially for the first excitation. In the case of a moving frame with $P=(1,0,0)$ the results are even inconsistent for a small basis and only become consistent with the mixed basis when the basis is enlarged. It should therefore be stressed that a suitable basis is crucial to extract physical results.

\subsection{Recent results for other QCD resonances}

While the extraction of the $\rho$ resonance is a great proof of principle that QCD calculations of resonance properties are feasible, this can only be the start. In this section recent results in other channels are reviewed.

\subsubsection{Meson-meson scattering in the $K\pi$, $D\pi$ and $D^\star\pi$ channels}

\begin{table}[b]
\begin{center}
\begin{tabular}{cccccccc}
$N_L^3\times N_T$ & $\kappa_l$ & $\beta$ & $a$[fm] & $L$[fm] & \#configs & $m_\pi$[MeV] & $m_K$[MeV]\\ 
\hline
$16^3\times32$ & 0.1283 & 7.1 & 0.1239(13) & 1.98 & 280/279 & 266(3)(3) & 552(2)(6)\\
\end{tabular}
\caption{Parameters for the nHYP-smeared Wilson-clover lattices \cite{Hasenfratz:2008ce,Hasenfratz:2008fg} used in \cite{Lang:2011mn,Lang:2012sv,Mohler:2012na}.}
\label{confs_anna}
\end{center}
\end{table}

Recently, first steps towards the determination of phase shifts and an extraction of resonance parameters have been taken for the case of meson-meson scattering in the $K\pi$, $D\pi$ and $D^\star\pi$ channels \cite{Lang:2012sv,Mohler:2012na}. For both of these studies configurations with $N_f=2$ flavors of nHYP smeared Wilson-clover quarks \cite{Hasenfratz:2008ce,Hasenfratz:2008fg} were used. Table \ref{confs_anna} shows the relevant parameters. The distillation method \cite{Peardon:2009gh,Morningstar:2011ka} was used to construct an interpolator basis of several $\bar{q}q$ and meson-meson interpolators. In both cases only the frame with total momentum zero was considered, which does not suffer from mixing of even and odd partial waves \cite{Fu:2011xz,Leskovec:2012gb,Doring:2012eu,Gockeler:2012yj}.

\begin{figure}[tb]
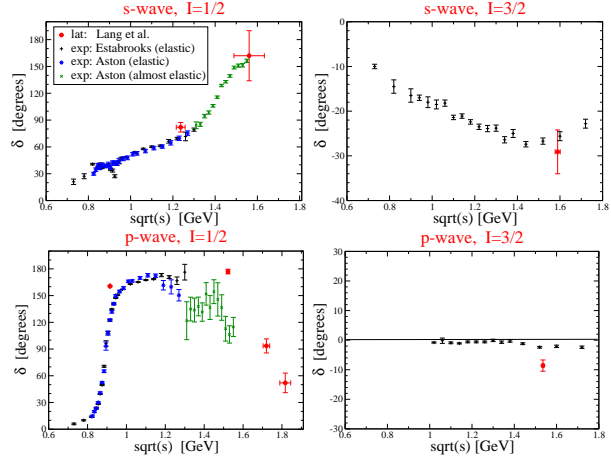

\begin{center}
\includegraphics[clip, height=3.0cm]{delta_s_half_sqrts_phys_lat12.eps}
\includegraphics[clip, height=3.0cm]{delta_s_threehalf_sqrts_phys_lat12.eps}\\
\includegraphics[clip, height=3.0cm]{delta_p_half_sqrts_phys_lat12.eps}
\includegraphics[clip, height=3.0cm]{delta_p_threehalf_sqrts_phys_lat12.eps}\\
\caption{Phase shift results for $K\pi$scattering in s-wave and p-wave for isospin $\frac{1}{2}$ and $\frac{3}{2}$.}
\label{delta_pionkaon}
\end{center}
\end{figure}

Figure \ref{delta_pionkaon} shows phase shift results from $K\pi$ scattering for isospin $\frac{1}{2}$ and $\frac{3}{2}$, in s-wave and p-wave. To illustrate the qualitative agreement, data extracted from experiment \cite{Estabrooks:1977xe,Aston:1987ir} is shown in black, blue and green while the lattice data is shown in red. For more information and some cautionary remarks please refer to \cite{Lang:2012sv}.

\begin{figure}[tb]
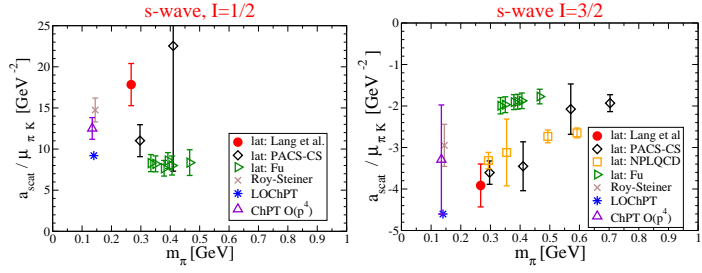

\begin{center}
\includegraphics[clip, height=3.5cm]{isospin_1-2_lat12.eps}
\includegraphics[clip, height=3.5cm]{isospin_3-2_lat12.eps}\\
\caption{Compilation of results for the isospin $\frac{1}{2}$ and $\frac{3}{2}$ s-wave $K\pi$ scattering lenghts taken from \cite{Lang:2012sv}. For more detail please refer to the text.}
\label{results_pionkaon}
\end{center}
\end{figure}

Figure \ref{results_pionkaon} shows recent lattice results \cite{Beane:2006gj,Sasaki:2009cz,Fu:2011wc,Lang:2012sv} for the s-wave scattering lenghts in units of the reduced mass $\mu_{k\pi}$\footnote{In leading order $\chi$PT this ratio does not depend on $m_\pi$.} compared to results from Chiral Perturbation Theory ($\chi$PT) \cite{Bernard:1990kw} or from a Roy-Steiner analysis \cite{Buettiker:2003pp}.

\begin{figure}[tb]
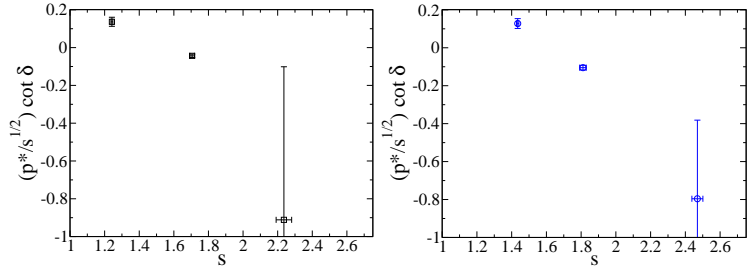

\begin{center}
\includegraphics[clip, height=3.5cm]{0+_kmatrix.eps}
\includegraphics[clip, height=3.5cm]{1+_kmatrix_modified.eps}
\caption{Phase shift results for the $D\pi$ and $D^\star\pi$ channels. To extract the data for $D^\star\pi$ scattering the heavy quark limit has to be assumed.}
\label{dmesons_both}
\end{center}
\end{figure}

A similar approach has been used to investigate $D\pi$ and $D^\star\pi$ scattering \cite{Mohler:2012na} where the charm quarks are treated with the Fermilab method \cite{ElKhadra:1996mp}. Figure \ref{dmesons_both} shows $\frac{p^\star}{\sqrt{s}}\cot{\delta_l}$ for the $D_0^\star$ and $D_1$ channels, since the combination  $\frac{p^*}{\sqrt{s}}\cot\delta_l=\frac{1}{g^2} (s-m_r^2)$ is linear in case of a Breit-Winger resonance. The left panel shows results in the $J^P=0^+$ channel where the $D_0^\star(2400)$ resonance is observed in experiment. Three levels have been extracted. 

In the $J^P=1^+$ channel there are two resonances $D_1(2420)$ and $D_1(2430)$ and without further assumptions the data is insufficient to extract resonance parameters. In the heavy quark limit, the narrow $D_1(2420)$ is expected to exclusively decay in d-wave, while the broad $D_1(2430)$ will decay exclusively in s-wave. In this limit, one then obtains the results displayed in the right panel of Figure \ref{dmesons_both}, where the level associated with the $D_1(2420)$ has been omitted.

\begin{figure}[tb] 
\begin{center}
\includegraphics[clip, height=4.5cm]{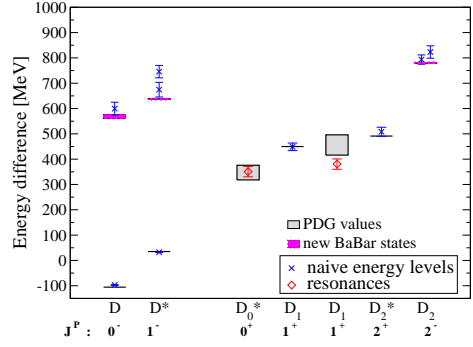}
\caption{Energy differences $\Delta E=E-\tfrac{1}{4}(M_{D}+3M_{D^*})$ for $D$ meson  states on the lattice and in experiment. Likely quantum numbers have been assigned to some of the states seen by BaBar \cite{delAmoSanchez:2010vq}. For more information please refer to the text and to \cite{Mohler:2012na}.}
\label{dmesons_results}
\end{center}
\end{figure}

\begin{table}[tbh]
\begin{center}
\begin{tabular}{ccc}
\hline
&$D_0^\star(2400)$&$D_1(2430)$\\
\hline
$g^{lat}$ [GeV] & $2.55 \pm 0.21$ & $2.01 \pm 0.15$\\
$g^{exp}$ [GeV] & $\le1.92 \pm 0.14$ & $\le2.50 \pm 0.40$\\
\hline
\end{tabular}
\caption{Resonance parameters for the $D_0^*(2400)$ and for the $D_1(2430)$ extracted in \cite{Mohler:2012na}. Experiment values are quoted as upper limits to reflect that only the total width is known.}
\label{dmesons_table}
\end{center}
\end{table}

Assuming a Breit-Wigner shape for both the $D_0^*(2400)$ and for the $D_1(2430)$, the resonance parameters displayed in Table \ref{dmesons_table} are obtained. In addition to the channels treated as resonances, energy levels were also extracted in other channels. The resulting spectrum is shown in Figure \ref{dmesons_results}. For further information including a detailed description of assumptions please refer to \cite{Mohler:2012na}. 

\subsubsection{$\kappa$ and $\sigma$ resonances from staggered simulations}

There are two recent papers \cite{Fu:2011xw,Fu:2012gf} aimed at the extraction of the $\kappa$ and $\sigma$ resonances using staggered quarks. In both cases the asqtad medium-coarse ensemble of size  $16^3\times48$ with $m_{u/d}=0.2m_s$ and $a\approx0.15$fm generated by the MILC collaboration has been used. As a basis one $\bar{q}q$ and one meson-meson interpolator, picking the goldstone pion/kaon $\pi_5$/$K_5$ is used. Extracting two energy levels and assuming a Breit-Wigner shape the resonance parameters in Table \ref{kappa_fu} are extracted. 

\begin{table}[tbh]
\begin{center}
\begin{tabular}{|c|c|}
\hline
$I=\frac{1}{2}\quad\pi K$&$I=0\quad\pi\pi$\\
\hline
$g_{\kappa\pi K}=4.54(76)$GeV & $g_{\sigma\pi\pi}=2.69(44)$GeV \\
$M_R=0.779(27)a$ & $M_R=0.691(37)a$\\
\hline
\end{tabular}
\caption{Resonance parameters attributed to the $\kappa$ and $\sigma$ resonances \cite{Fu:2011xw,Fu:2012gf}. For cautionary remarks please refer to the text.}
\label{kappa_fu}
\end{center}
\end{table}

It should however be pointed out that the $\bar{q}q$ interpolators inevitably couple to all staggered taste combinations \cite{Prelovsek:2005rf,Bernard:2007qf}. Therefore, the variational analysis may render other taste combinations $K_b\pi_b$ and $\pi_b\pi_b$ as excited states. Moreover the experimental data already shown in Figure \ref{delta_pionkaon} for the $\kappa$ channel makes the assumption of a Breit-Wigner shape questionable.

\subsubsection{$\Delta(1232)\leftrightarrow N\pi$ by the QCDSF collaboration}

\begin{figure}[tb]
\begin{center}
\includegraphics[clip, height=4.0cm]{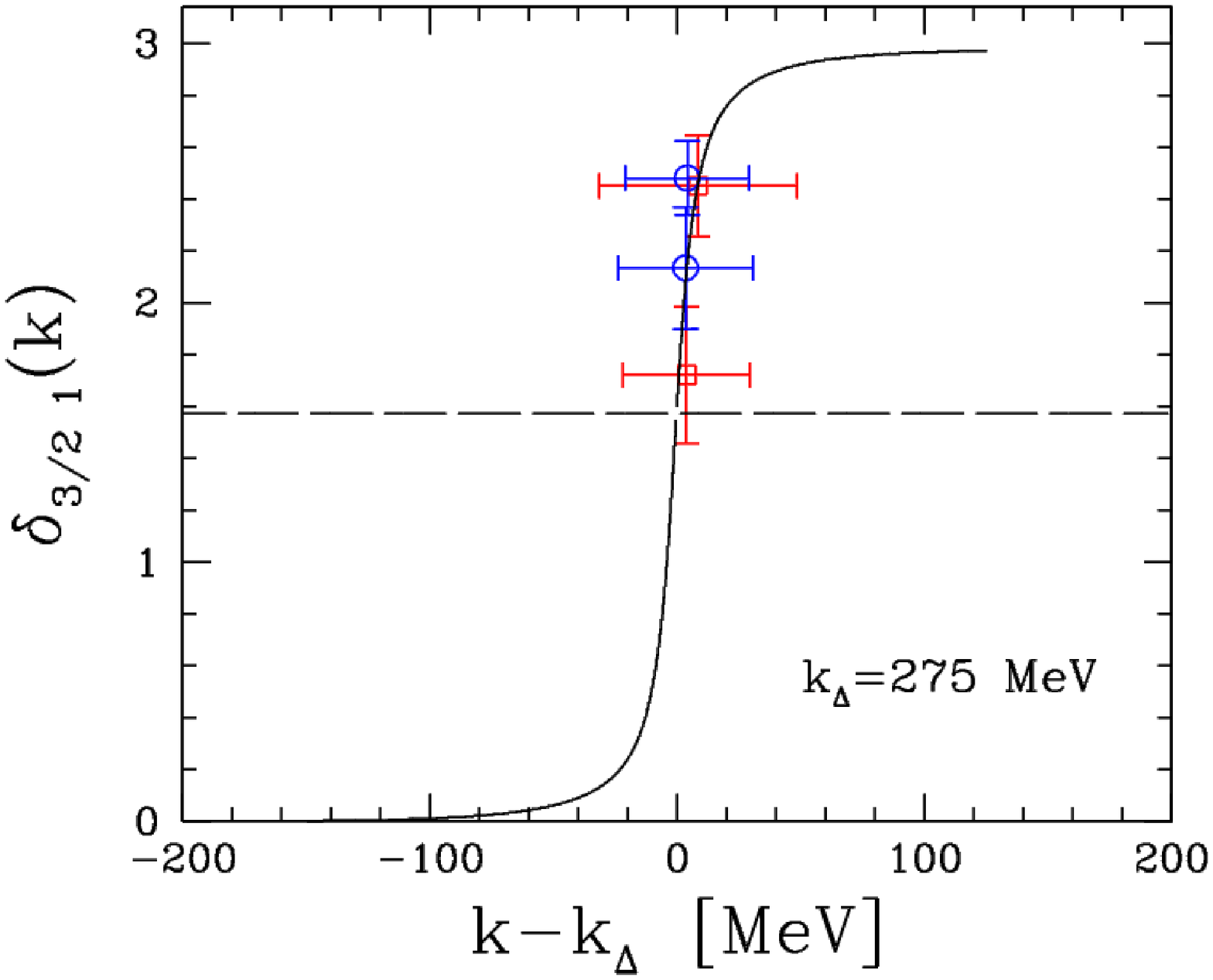}
\includegraphics[clip, height=4.0cm]{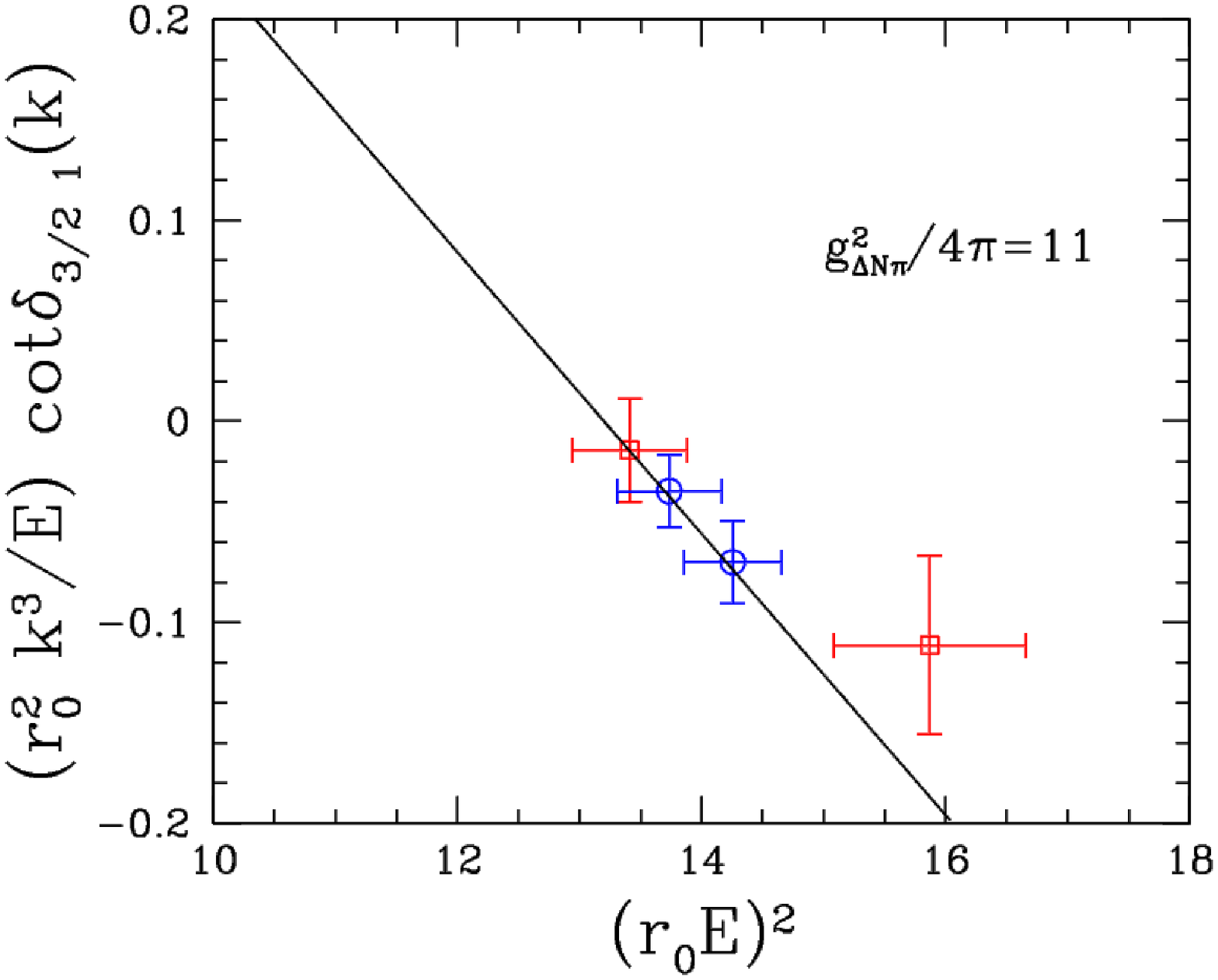}
\caption{Preliminary results from a simulation of nucleon-pion scattering by the QCDSF collaboration \cite{qcdsf_private}. The plot shows data in the vicinity of the $\Delta(1232)$ resonance.}
\label{delta_QCDSF}
\end{center}
\end{figure}

The QCDSF collaboration is investigating nucleon-pion scattering in the $\Delta(1232)$ channel \cite{qcdsf_private} using $32^3\times T$, $40^3\times T$ and $48^3\times T$ lattices with a pion mass $m_\pi\approx250$ MeV. Preliminary results for the phase shift are plotted in Figure \ref{delta_QCDSF}. For the delta-pion-nucleon coupling they obtain $\frac{g_{\Delta\pi N}^2}{4\pi}=11\pm_3^4$ which should be compared to the value $\frac{g_{\Delta\pi N, exp}^2}{4\pi}\approx 14.4$ extracted from experiment. 

\subsubsection{Baryon interactions in a matrix Hamiltonian model}

\begin{figure}[tb]
\begin{center}
\includegraphics[clip, height=4.0cm]{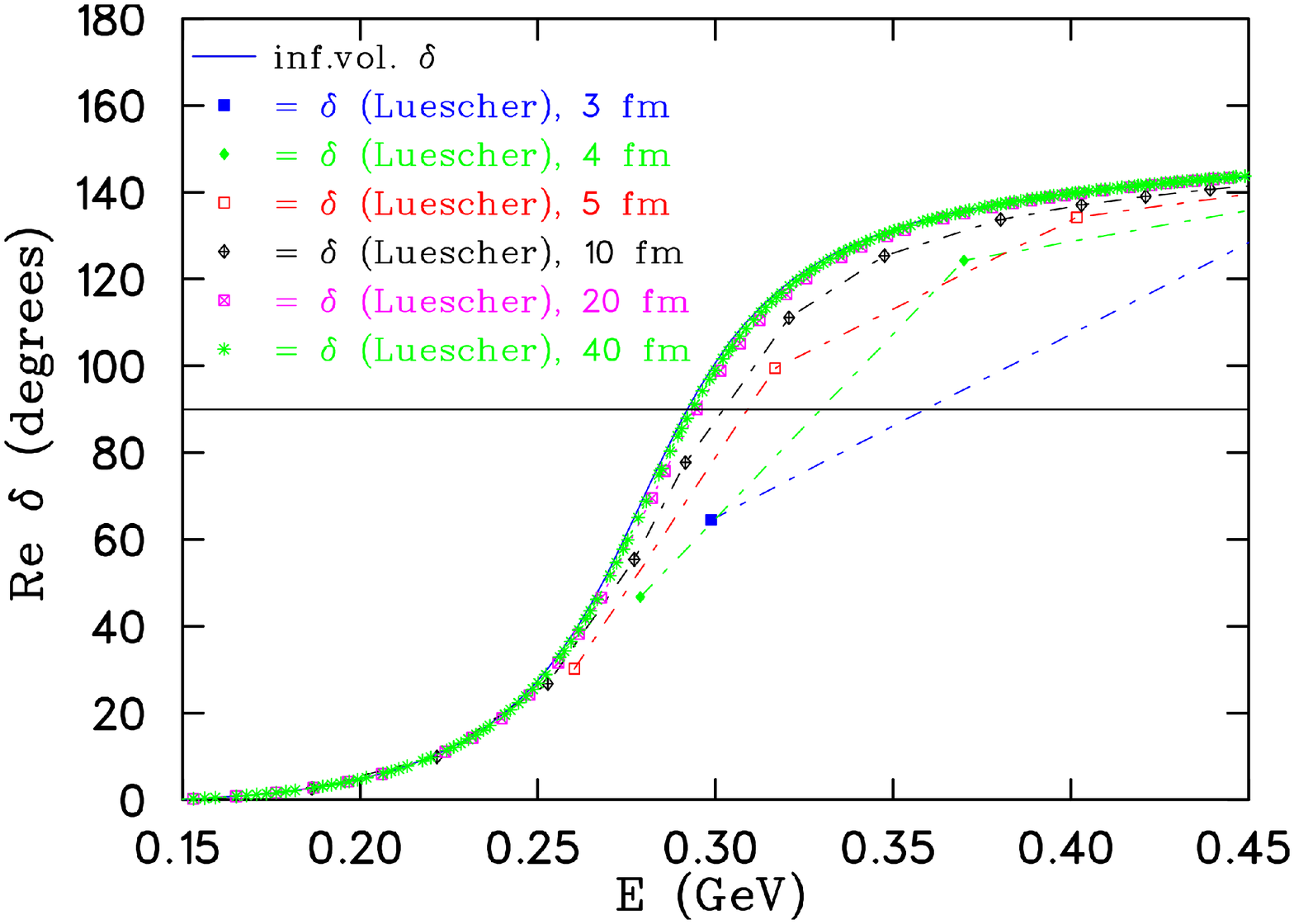}
\includegraphics[clip, height=4.0cm]{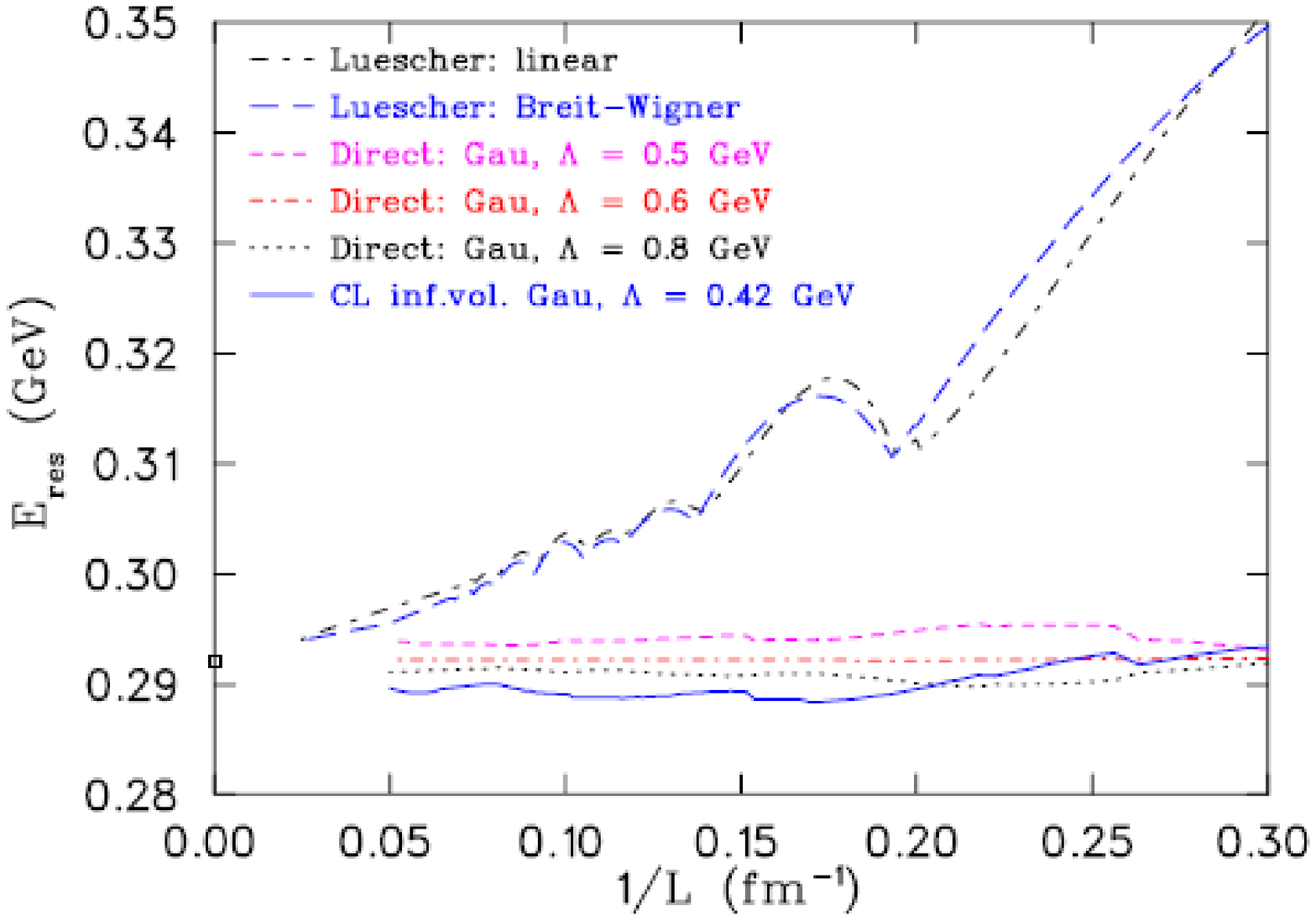}
\caption{Left panel: Estimates of the phase shift employing L\"uscher's formula to the $\Delta N\pi$ model for different lattice extent $L$; Right panel: The resonance energy from the L\"uscher method compared to a direct calculation. Plots taken from \cite{Hall:2012wz}.}
\label{baryon_int}
\end{center}
\end{figure}

 In L\"uscher's method finite volume errors should be exponentially suppressed as a function of the lattice size $L$. This assumes that hadrons at the boundary of the box are in the asymptotic region, otherwise the error may scale as $L^{-1}$. In \cite{Hall:2012wz} this issue is investigated  in a matrix Hamiltonian model for $\Delta\leftrightarrow N\pi$. Figure \ref{baryon_int} shows some of the results. In the left panel results from the L\"uscher method using a different spatial extent are plotted. In the right panel the results from the L\"uscher method are compared to the direct calculation. While both methods agree for large volumes,  it is clear that large finite volume effects are present in this case. For further discussion please refer to \cite{Hall:2012wz}.

\section{Beyond QCD}

As already demonstrated in the case of toy models, nonperturbative methods for the extraction of resonances in a lattice simulation can also be applied to problems beyond QCD. As an example, the L\"uscher method has been used recently in a simulation of the pure Higgs-Yukawa sector of the electroweak standard model \cite{Gerhold:2011mx}. The model consist of a complex scalar Higgs doublet and a mass-degenerate fermion doublet (representing top and bottom quarks) coupled in a chirally invariant way. In this case properties of the Higgs resonance can be extracted.

\begin{figure}[tb]
\begin{center}
\includegraphics[clip, height=3.7cm]{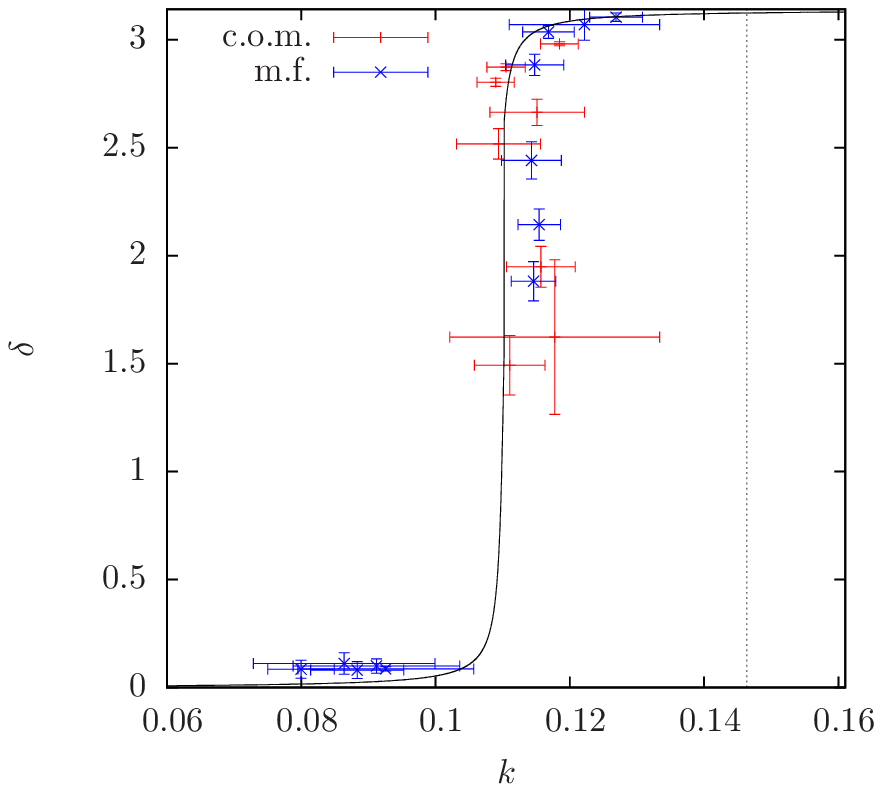}
\includegraphics[clip, height=3.7cm]{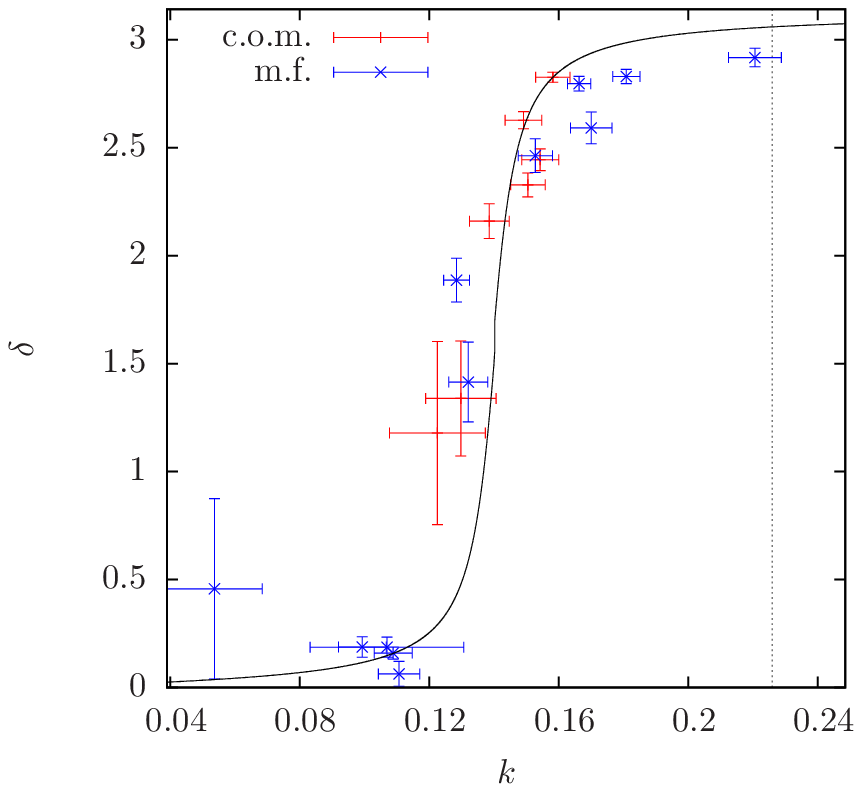}
\includegraphics[clip, height=3.7cm]{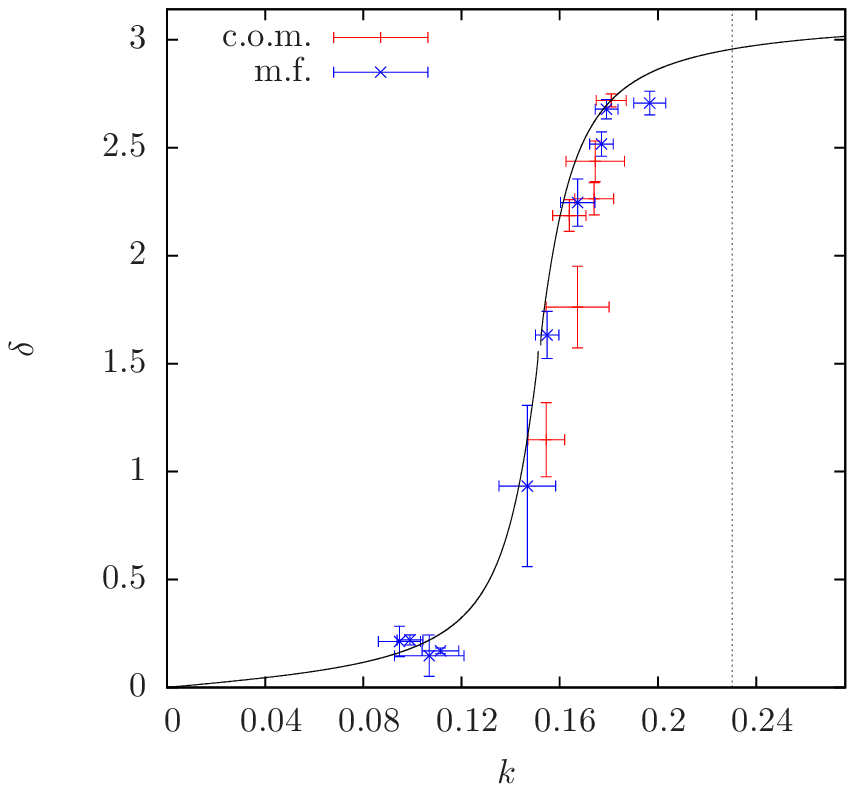}
\caption{Elastic scattering phases obtained for three different quartic couplings $\hat{\lambda}$ within a chiral Higgs-Yukawa model. Plots taken from \cite{Gerhold:2011mx}.}
\label{higgs_model}
\end{center}
\end{figure}

Figure \ref{higgs_model} shows the results for three different quartic couplings $\hat{\lambda}$. While this model is quite distinct from the situation in nature, is is very interesting to observe that the width of the Higgs stays small even at large coupling.

\section{Summary \& Outlook}

While the field is progressing fast and the last year has seen first QCD resonance studies in channels other than the $\rho$ meson channel, lattice studies of (QCD) resonances are still in their infancy. Part of the reason is that the standard L\"uscher method is limited to the case of elastic scattering. While there has been some progress recently \cite{He:2005ey,Doring:2011vk,Briceno:2012yi,Hansen:2012tf,Guo:2012hv} most ideas to go beyond the case of elastic scattering require a certain degree of modeling.

To illustrate the severity of this limitation it is instructive to appeal to experiment once more. Even in the low-lying meson spectrum, there are several interesting cases where multiple thresholds are expected to be important. Two prominent examples would be the $a_0(980)$ which goes both to $\eta\pi$ and $\bar{K}K$ and the $K_1(1270)$ where the branching ratios to $K\rho$, $K^\star(872)\pi$, $K\omega$ and $K_0^\star(1430)\pi$ are all known to be sizable.

 Last but not least there are also many interesting states, especially the so-called X, Y and Z states in the charmonium spectrum, where a first principle calculation from lattice QCD is needed, but where most interesting states are above multiparticle thresholds. In short, there is much progress to be made for a comprehensive description of QCD resonances, and beyond.

\acknowledgments
I would like to thank the organizers for the invitation to present this review. In addition I thank Andrei Alexandru, Albert Deuzeman, Christian B.~Lang, Craig Pelissier, Sasa Prelovsek and Gerrit Schierholz for providing me with material/information for the talk. Finally, I would like to thank my collaborators Christian B.~Lang, Luka Leskovec, Sasa Prelovsek and Richard Woloshyn. Fermilab is operated by Fermi Research Alliance, LLC under Contract No. De-AC02-07CH11359 with the United States Department of Energy.

\providecommand{\href}[2]{#2}\begingroup\raggedright\endgroup

\end{document}